\input harvmac\skip0=\baselineskip
\input epsf

\newcount\figno
\figno=0
\def\fig#1#2#3{
\par\begingroup\parindent=0pt\leftskip=1cm\rightskip=1cm\parindent=0pt
\baselineskip=11pt \global\advance\figno by 1 \midinsert
\epsfxsize=#3 \centerline{\epsfbox{#2}} \vskip 12pt {\bf Fig.\
\the\figno: } #1\par
\endinsert\endgroup\par
}
\def\figlabel#1{\xdef#1{\the\figno}}
\def\encadremath#1{\vbox{\hrule\hbox{\vrule\kern8pt\vbox{\kern8pt
\hbox{$\displaystyle #1$}\kern8pt} \kern8pt\vrule}\hrule}}

\def\o{\omega }

\def\a{{\alpha}}
\def\b{{\beta}}

\def\msurr{\mathsurround=0pt}
\def\overleftrightarrow#1{\vbox{\msurr\ialign{##\crcr
        $\leftrightarrow$\crcr\noalign{\kern-1pt\nointerlineskip}
        $\hfil\displaystyle{#1}\hfil$\crcr}}}
\noblackbox

\lref\KlebanovReview{I.~R.~Klebanov, ``String Theory in
Two-Dimensions", hep-th/9108019.}

\lref\GandM{P.~Ginsparg and G.~W.~Moore, ``Lectures on 2-D gravity
and 2-D String Theory", hep-th/9304011.}

\lref\PolchinskiReview{J.~Polchinski, ``What is String Theory?",
hep-th/9411028.}

\lref\TadashiNick{T.~Takayanagi and N.~Toumbas, ``A Matrix Model
Dual of Type 0B String Theory in Two Dimensions", hep-th/0306177.}

\lref\NewHat{M.~R.~Douglas, I.~R.~Klebanov, D.~Kutasov,
J.~Maldacena, E.~Martinec and N.~Seiberg, ``A New Hat for the
$c=1$ Matrix Model", hep-th/0307195.}

\lref\MPR{G.~W.~Moore, M.~R.~Plesser and S.~Ramgoolam, ``Exact
S-matrix for 2-D String Theory", hep-th/9111035.}

\lref\DMR{R.~Dijkgraaf, G.~W.~Moore and R.~Plesser, ``The
Partition Function of 2-D String Theory", hep-th/9208031.}

\lref\StringEq{I.~Kostov, ``String Equation for String Theory on a
Circle", hep-th/0107247.}

\lref\UpsideDown{D.~Boulatov and V.~Kazakov, ``One-Dimensional
String Theory with Vortices as Upside-Down Matrix Oscillator",
hep-th/0012228.}

\lref\KKK{V.~Kazakov, I.~Kostov and D.~Kutasov, ``A Matrix Model
for the Two Dimensional Black Hole", hep-th/0101011.}

\lref\AKKtime{S.~Alexandrov, V.~Kazakov and I.~Kostov,
``Time-Dependent Backgrounds of 2D String Theory",
hep-th/0205079.}

\lref\AKKnormal{S.~Alexandrov, V.~Kazakov and I.~Kostov, ``2D
String Theory as Normal Matrix Model", hep-th/0302106.}

\lref\DRSVW{ O.~DeWolfe, R.~Roiban, M.~Spradlin, A.~Volovich and
J.~Walcher, ``On the S-matrix of Type 0 String Theory",
hep-th/0309148.}

\lref\Reloaded{J.~McGreevy and H.~Verlinde, ``Strings from
Tachyons: The $c=1$ Matrix Reloaded", hep-th/0304224}

\lref\CQDbrane{J.~McGreevy, J.~Teschner and H.~Verlinde,
``Classical and Quantum D-branes in 2D String Theory",
hep-th/0305194.}

\lref\KMSdecay{I.~R.~Klebanov, J.~Maldacena and N.~Seiberg,
``D-brane Decay in Two-Dimensional String Theory",
hep-th/0305159.}

\lref\tadashi{A.~Strominger and T.~Takayanagi, unpublished notes.}

\lref\MSY{A.~Maloney, A.~Strominger and X.~Yin, ``S-brane
Therodyamics", hep-th/0302146.}

\lref\KostovReview{I.~Kostov, ``Integrable Flows in $c=1$ String
Theory", hep-th/0208034.}

\lref\imagineryD{D.~Gaiotto, N.~Itzhaki and L.~Rastelli, ``Closed
Strings as Imaginery D-branes", hep-th/0304192.}

\lref\CKLM{C.~G.~Callan, I.~R.~Klebanov, A.~W.~W.~Ludwig,
J.~M.~Maldacena, ``Exact Solution of a Boundary Conformal Field
Theory", hep-th/9402113.}

\lref\Jevicki{A.~Jevicki, ``Developments in 2D String Theory",
hep-th/9309115.}

\lref\fermiBH{A.~Giveon, A.~Konechny, A.~Pakman, A.~Sever, ``Type
0 Strings in a 2-d Black Hole", hep-th/0309056. }

\lref\Sen{A.~Sen, ``Rolling Tachyon", hep-th/0203211; ``Tachyon
Matter", hep-th/0203265; ``Field Theory of Tachyon Matter",
hep-th/0204143.}

\lref\CSmm{M.~Aganagic, A.~Klemm, M.~Marino, C.~Vafa, ``Matrix
Model as a Mirror of Chern-Simons Theory", hep-th/0211098.}

\lref\UenoTakasaki{K.~Ueno and K.~Takasaki, ``Toda lattice
hierarchy", in {\sl Advanced Studies in Pure Mathematics 4}, ed.
K.~Okamoto, North-Holland, Amsterdam, 1-94 (1984).}

\lref\Hirota{R.~Hirota, ``Direct Method in Soliton Theory", {\sl
Solitons}, ed. R.~K.~Bullogh and R.~J.~Caudrey, Springer (1980).}

\lref\Johnson{C.~Johnson, ``Non-Perturbative String Equations for
Type 0A", hep-th/0311129.}

\lref\MatrixCosmology{J.~L.~Karczmarek and A.~Strominger, ``Matrix
Cosmology", hep-th/0309138. }

\lref\LLM{N.~Lambert, H.~Liu and J.~Maldacena, ``Closed Strings
from Decaying D-branes", hep-th/0303139.}

\lref\Mukhi{S.~Mukhi, ``Topological Matrix Models, Liouville
Matrix Model and $c=1$ String Theory", hep-th/0310287.}

\lref\Taylor{W.~Taylor, ``D-brane Field Theory on Compact Spaces",
hep-th/9611042.}

\lref\IKKT{N.~Ishibashi, H.~Kawai, Y.~Kitazawa and A.~Tsuchiya,
``A Large N Reduced Model as Superstring", hep-th/9612115. }

\lref\AST{T.~Asakawa, S.~Sugimoto and S.~Terashima, ``D-branes,
Matrix Theory and K-homology", hep-th/0108085.}

\lref\BHOA{S.~Gukov, T.~Takayanagi and N.~Toumbas, ``Flux
Backgrounds in 2D String Theory", hep-th/0312208.}

\lref\Gross{D.~Gross and J.~Walcher, ``Non-perturbative RR
Potentials in the c=1 Matrix Model", hep-th/0312021.}

\lref\ZZsupera{T.~Fukuda and K.~Hoscmichi, ``Super Liouville
Theory with Boundary", hep-th/0202032. }

\lref\Hori{K.~Hori and A.~Kapustin, ``Duality of the Fermionic 2d
Black Hole and N=2 Liouville Theory as Mirror Symmetry",
hep-th/0104202.}

\lref\ZZsuperb{C.~Ahn, C.~Rim and M.~Stanishkov, ``Exact One-Point
Function of N=1 super-Liouville Theory with Boundary",
hep-th/0202043.}

\lref\CIS{O.~Babelon, D.~Bernard and M.~Talon, {\sl Introduction
to Classical Integrable Systems}, Cambridge University Press.}

\lref\GrossWitten{D.~Gross and E.~Witten, ``Possible Third Order
Phase Transition in the Large N Lattice Gauge Theory",
Phys.Rev.{\bf D21}, 446 (1980).}

\lref\postNewHat{I.~R.~Klebanov, J.~Maldacena and N.~Seiberg,
``Unitary and Complex Matrix Models as 1-d Type 0 Strings",
hep-th/0309168.}

\lref\Kutasov{A.~Giveon and D.~Kutasov, ``Little String Theory in
a Double Scaling Limit", hep-th/9909110.}

\lref\Kutasovb{A.~Giveon and D.~Kutasov, ``Comments on Double
Scaled Little String Theory", hep-th/9911039.}

\lref\Kutasovc{D.~Kutasov and D.~A.~Sahakyan, ``Comments on the
Thermodynamics of Little String Theory", hep-th/0012258.}

\lref\AKKnp{S.~Yu.~Alexandrov, V.~Kazakov and D.~Kutasov,
``Non-Perturbative Effects in Matrix Models and D-branes",
hep-th/0306177.}

\lref\AKthermal{S.~Yu.~Alexandrov and V. Kazakov, ``Thermodynamics
of 2D string theory", hep-th/0210251.}

\lref\Tseytlin{V.~Kazakov and A.~Tseytlin, ``On free energy of 2-d
black hole in bosonic string theory", hep-th/0104138.}

\lref\ADStwo{A.~Strominger, ``A Matrix Model for AdS2",
hep-th/0312194.}

\lref\Schomerus{S.~Ribault and V.~Schomerus, ``Branes in the 2D
black hole", hep-th/0310024.}

\lref\Dasguptaa{S.~Dasgupta and T.~Dasgupta, ``Renormalization
Group Approach to $c=1$ Matrix Model on a circle and D-brane
Decay", hep-th/0310106.}

\lref\Dasguptab{S.~Dasgupta and T.~Dasgupta, ``Nonsinglet Sector
of $c=1$ Matrix Model and 2D Black Hole", hep-th/0311177.}

\lref\Murthy{J.~McGreevy, S.~Murthy and H.~Verlinde,
``Two-dimensional superstrings and the supersymmetric matrix
model", hep-th/0308105.}

\lref\ZZ{A.~B.~Zamolodchikov and Al.~B.~Zamolodchikov, ``Liouville
field theory on a pseudosphere", hep-th/0101152.}

%
%

\Title{\vbox{\baselineskip12pt\hbox{hep-th/0312236}
}}{\vbox{\centerline{Matrix Models, Integrable
Structures,}\bigskip \centerline{and T-duality of Type 0 String
Theory}}}

\centerline{Xi Yin}
\smallskip
\centerline{ Jefferson Physical Laboratory} \centerline{Harvard
University, Cambridge, MA } \vskip .3in \centerline{\bf Abstract}
{ Instanton matrix models (IMM) for two dimensional  string
theories are obtained from the matrix quantum mechanics (MQM) of
the T-dual theory. In this paper we study the connection between
the IMM and MQM, which amounts to understand T-duality from the
viewpoint of matrix models. We show that type 0A and type 0B
matrix models perturbed by purely closed string momentum modes (or
purely winding modes) have the integrable structure of Toda
hierarchies, extending the well known results for $c=1$ string. In
particular, we show that type 0A(0B) MQM perturbed by momentum
modes has the same integrable structure as type 0B(0A) MQM
perturbed by winding modes, which is a nontrivial check of the
T-duality between the matrix models. The MQM deformed by NS-NS
winding modes are used to study type 0 string in 2D black holes.
We also find an intriguing connection between the IMM and the MQM
via tachyon condensation. The array of alternating D-instantons
and anti-D-instantons separated at the critical distance plays a
key role in this picture. We discuss its implications on sD-branes
in two dimensional string theories. }

\smallskip
\Date{}

\listtoc
\writetoc

\newsec{Introduction }
Recently the $c=1$ matrix quantum mechanics(for reviews see
\refs{\KlebanovReview,\GandM,\PolchinskiReview}) has received a
lot of attention because of its new interpretation as the
decoupled world volume theory of unstable
D0-branes\refs{\Reloaded,\CQDbrane,\KMSdecay}. The matrix models
dual to type 0 string theories were also proposed in
\refs{\TadashiNick,\NewHat}. For other recent developments, see
\refs{\MatrixCosmology, \Murthy, \Johnson, \BHOA, \Dasguptaa,
\Dasguptab, \Schomerus, \Gross, \ADStwo}. The type 0B matrix
quantum mechanics (MQM) describes open string tachyons living on
the unstable D0-branes, whereas the type 0A MQM describes
tachyonic open strings stretched between stable D0- and
anti-D0-branes. Upon compactification on Euclidean time, these two
matrix models are conjectured to be T-dual to each other. The
exact agreement in free energy was found in \NewHat. However,
unlike $c=1$ matrix model which can be derived from discretizing
the Polakov action on the string world sheet, such a derivation is
not known for type 0 matrix models. From the matrix model point of
view, the T-duality between type 0A and 0B strings seems rather
mysterious. To understand this T-duality is one of the motivations
of this paper.

To start, let us consider type 0A MQM with Euclidean time
compactified on a circle of radius $\alpha'/R$. This is supposed
to be T-dual to type 0B string theory on a circle of radius $R$.
By decomposing the fields in type 0A MQM into their Fourier modes
along the thermal circle, we get a zero-dimensional matrix model,
of the form \eqn\zmm{ \int dU d\tilde U \prod_n dt_n dt_n^\dagger
e^{-\tilde\beta{\rm Tr}\left[(Xt_n-t_n\tilde X+2\pi
nRt_n)(Xt_n-t_n\tilde X+2\pi
nRt_n)^\dagger-a^2t_nt_n^\dagger\right] } } where $U=e^{iX/R}$,
$\tilde U=e^{i\tilde X/R}$ are the holonomies of the Wilson lines
in type 0A MQM, $t_n,t_n^\dagger$ are the Fourier modes of the
complex tachyons. This is an instanton matrix model (IMM), since
the Wilson lines on the D0-branes in type 0A theory are mapped to
collective coordinates $X,\tilde X$ for D-instantons in type 0B
theory, and the tachyons on the D0-branes are mapped to tachyonic
open strings stretched between D-instantons and anti-D-instantons.
Similar D-instanton matrix models have been studied in the context
of ten dimensional type IIB string theory\refs{\IKKT,\AST}. Now
having two different matrix model duals of type 0B theory, we want
to understand how they are related to each other. Since one of
them involves unstable D0-branes, and the other involves D- and
anti-D-instantons, it is natural to suspect that tachyon
condensation plays the key role of connecting the two theories. If
we can show the equivalence between the IMM and MQM, we would
prove the T-duality between type 0A and 0B matrix models.

The first thing we want to understand is the identification
between operators in IMM and in MQM, in particular the ones that
correspond to closed string momentum and winding modes. This was
well understood in the context of $c=1$ string theory. The
momentum modes in MQM are represented as asymptotic perturbations
of the fermi sea, whereas the winding modes are identified with
the holonomy of the gauge fields\refs{\UpsideDown,\KKK}. $c=1$
string perturbed purely by momentum modes\refs{\DMR,\AKKtime}, or
by winding modes\refs{\KKK}, has the integrable structure of Toda
lattice hierarchy, with the integrable flow generated by the
corresponding closed string perturbations. In particular, the
perturbed grand canonical partition function is shown to be the
$\tau$-function of the corresponding integrable hierarchy. The
integrable structure appearing in $c=1$ string theory is subject
to constraints, known as the string equation\refs{\StringEq}.
These constraints can be equivalently thought of as imposing an
initial condition on the flow of $\tau$-functions, which is the
unperturbed partition function. Since the unperturbed $c=1$ MQM on
radius $R$ and $\alpha'/R$ have the same free energy, it follows
from the integrable structures that the grand partition function
of two theories perturbed respectively by momentum modes and
winding modes also agree.

We shall generalize this approach to type 0 string theories. In
type 0B MQM, the symmetric and antisymmetric perturbations of the
fermi sea decouple.\foot{As remarked in \DRSVW, this doesn't mean
that the NS-NS and R-R scalars decouple, because of the
nonlinearity of bosonization.} They generate two independent Toda
flows, subject to different string equations. Consequently the
perturbed partition function is the product of two
$\tau$-functions, associated with the symmetric and antisymmetric
perturbations respectively. On the type 0B IMM side, one can
integrate out the tachyons and get a unitary matrix model only in
terms of the ``holonomies", or the collective coordinates of the
instantons. This is T-dual to the ``twisted partition" discussed
in \refs{\UpsideDown,\KKK} in the case of $c=1$ string. As we will
see, the perturbed grand canonical partition function of the IMM
indeed has a similar structure, provided a nontrivial
identification of the perturbation parameters with those of MQM.
We will use it to identify the operators in IMM that are dual to
NS-NS and R-R scalars in spacetime.

Type 0A MQM and IMM perturbed by momentum modes also have the
structure of Toda hierarchy. This is very similar to the case of
$c=1$ string, except that the string equations for type 0A MQM and
IMM are nonperturbatively well defined. The combination of these
shows that both type 0A and type 0B theories perturbed purely by
momentum modes, or purely by winding modes, are integrable. This
also directly verifies that the IMM and MQM are equivalent at
least when only momentum modes or winding modes are present. The
perturbations involving both momentum and winding modes are more
complicated, since in that case we would lose integrability.

An alternative attempt to connect the IMM with MQM, which is more
direct and intuitive, is via tachyon condensation. Consider
turning on an open string tachyon profile on the D0-brane world
volume $T(X)\sim \lambda\cos (X/\sqrt{2\alpha'})$, where $X$ is
the Euclidean time. As well known \refs{\CKLM,\Sen} this
corresponds to a marginal deformation in the worldsheet CFT. With
sufficiently large $\lambda$, it takes a D0-brane into an array of
alternating D- and anti-D-instantons separated at the critical
distance. It is natural to expect that, the MQM expanded near this
tachyon profile should be the same as the IMM expanded near the
configuration of an array of D-instantons.

On the MQM side, this periodic tachyon profile effectively
discretizes the Euclidean time circle to a periodic lattice of
spacing $\pi\sqrt{2\alpha'}$. It is well known\KlebanovReview\
that the MQM on a time lattice of spacing $\epsilon$ with
$\epsilon$ less than a critical distance $a$ is exactly equivalent
to the continuum theory, provided proper redefinition of the
parameters. In our case $a=\pi\sqrt{2\alpha'}$ is the same
critical distance at which the tachyonic open string stretched
between D- and anti-D-instantons becomes massless. The array of
D-instantons is classically a stationary configuration in the IMM.
If we integrate out the (complex) tachyons connecting the
instantons, we expect an instability that drives the D- and
anti-D-instantons toward each other. On the other hand, the
collective coordinates of the instantons are eigenvalues of
Hermitian matrices in the IMM. They effectively repel each other
and fill up a ``sea" of D-instantons. We have a large number of
D-instantons distributed in a periodic effective potential $V(X)$
along the thermal circle, and the D-instanton array corresponds to
a critical point where the instantons are sitting at the top of
the potential. There is a phase transition when the ``instanton
sea" merges the top of the potential, and {\sl this is the
critical point that defines the double scaling limit of the IMM}.
This is analogous to the case of $c=1$ matrix model, where the
double scaling limit is defined as the limit that the fermi level
approaches the top of the tachyon potential. By relating the
collective coordinates of the array of D-instantons to the open
string tachyons in the discretized MQM at the sites on the time
lattice, we will show that the two matrix models become the same
in the limit of critical distance.

\fig{Schematic picture of the array of D-instantons condensing
along the Euclidean time circle into the ``instanton
sea".}{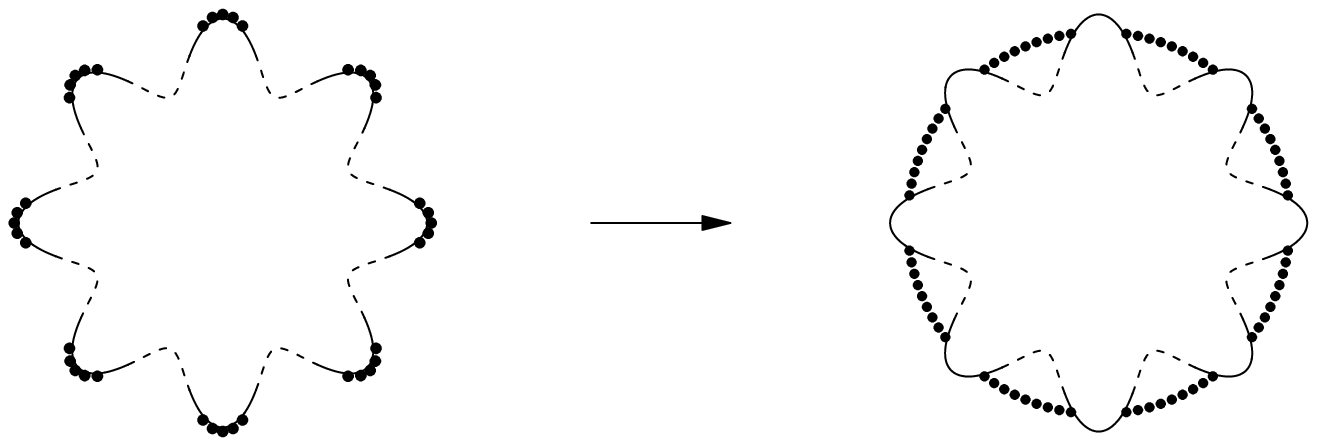}{5.0truein}

It was proposed in \MSY\ (see also \refs{\LLM,\imagineryD}) that
sD-branes ($\lambda=1/2$ s-brane) are described by an array of D-
and anti-D-instantons along the Euclidean time, separated at the
critical distance (see also \imagineryD). In the sense described
above, the IMM can be thought of as the world volume theory of
sD-branes, and sD-branes play the same role in the IMM as
D0-branes in the MQM. We compute the closed string fields sourced
by sD-branes from IMM, and find exact agreement with calculation
using ZZ boundary states. In fact, we can reproduce the $(1,1)$ ZZ
boundary state in Liouville \refs{\ZZ} and super-Liouville theory
\refs{\ZZsupera,\ZZsuperb} from $c=1$ IMM and type 0B IMM in a
very simple manner.

The $c=1$ matrix model deformed by the lowest winding modes is
conjectured to describe bosonic string in a 2D Euclidean black
hole \KKK. The conjecture is extended to type 0 theory in
\fermiBH. The exact free energy of the deformed type 0 matrix
models can in principal be obtained by solving the Hirota
differential equations. Using the technique of \KKK, we can solve
for the genus 0 and genus 1 terms in the perturbative expansion of
the free energy. They are of the form\refs{\KKK,\Tseytlin}
\eqn\freeintro{ {\cal F} = -2\pi (R-R_H) M +(s_1+1) \ln M +\cdots
} where $M$, depending on the coefficient $\lambda$ of the winding
mode deformation, is interpreted as the mass of the black hole,
and $R_H$ is the asymptotic radius in Euclidean time. In the
T-dual ${\cal N}=2$ Liouville theory \refs{\Kutasov,\Hori},
$\lambda$ is essentially the coefficient of the Liouville term. We
find from the genus 0 piece of the free energy that $M\propto
\lambda^4$, which agrees with the general behavior expected from
${\cal N}=2$ Liouville theory. The density of states for the black
hole has Hagedorn growth \eqn\densintro{ \rho(M)\sim M^{s_1}
e^{\beta_H M} } We find that $s_1=-17/12$ for type 0B string and
$s_1=-13/12$ for type 0A string (uncharged black hole).

This paper is organized as follows. In section 2 we derive the IMM
for $c=1$ and type 0 strings. In section 3, we review the
integrable structure in the S-matrix of $c=1$ MQM, and extend them
to the case of type 0A and type 0B MQM. We show that the same
integrable structures appear in the corresponding IMMs, and that
they are subject to the same string equations by computing the
unperturbed free energies. Section 4 studies the connection
between IMM and MQM from the viewpoint of tachyon condensation. In
section 5, we compute the closed string fields sourced by
sD-branes from IMM and compare them to the results obtained from
ZZ boundary states. The integrable structures of type 0 theories
are applied to computing the free energy of the 2D Euclidean black
hole in section 6.

\newsec{The Instanton Matrix Model}

In this section we derive the instanton matrix model of $c=1$
string and type 0 string theories from the (gauged) matrix quantum
mechanics of the T-dual theory. We will set $\alpha'=1$ in the
case of $c=1$ string, and keep $\alpha'$ explicitly in most of the
discussions on type 0 theories, unless otherwise indicated.

\subsec{The IMM for $c=1$ string}

Consider the gauged $c=1$ MQM with Euclidean time $x$ compactified
on a circle of radius $R'=1/R$. The matrix model action is
\eqn\caact{ S = \int_0^{2\pi/R}dx {\rm Tr}\left[ {1\over
2}(\partial_x\Phi+i[A,\Phi])^2 -{1\over 2}\Phi^2 \right] } where
the tachyon $\Phi(x)$ and the gauged field $A(x)$ are Hermitian
$N\times N$ matrices. Let us fix the gauge $\partial_xA=0$, which
sets $A$ to its zero mode $A^{(0)}$ and introduces the
Fadeev-Popov determinant \eqn\ghostdetca{ \int db\,dc\exp({\rm
Tr}b\partial_xD_xc) = \prod_{i<j}\left( {\sin[(a_i-a_j)/2R]\over
(a_i-a_j)/2R} \right)^2 } where $a_i$'s are the eigenvalues of
$A^{(0)}$. Let us decompose $\Phi(x)$ into its momentum modes
$\Phi^{(n)}$ along the thermal circle. The action is now written
as \eqn\immactca{ \eqalign{ S = {2\pi\over R} \sum_{n\in{\bf Z}}
{\rm Tr} \left[ {1\over 2}\left(nR
\Phi^{(n)}+[A^{(0)},\Phi^{(n)}]\right) \left(nR
\Phi^{(-n)}-[A^{(0)},\Phi^{(-n)}]\right)
-{1\over2}\Phi^{(n)}\Phi^{(-n)} \right] } } In the T-dual theory
$X\equiv 2\pi A^{(0)}$ is the collective coordinate for
D-instantons, $\Phi^{(n)}$ are the open string tachyons stretched
between D-instantons with (relative) winding number $n$. T-duality
acts on the inverse ``Planck constant" $\beta$ as $\beta\to\beta
R$. We end up with an instanton matrix model action \eqn\immactcb{
\eqalign{ S = {\beta\over 2\pi} \sum_{n\in{\bf Z}} {\rm Tr} \left[
{1\over 2}\left(2\pi nR \Phi^{(n)}+[X,\Phi^{(n)}]\right)
\left(2\pi nR \Phi^{(-n)}-[X,\Phi^{(-n)}]\right) \right.\cr\left.
-{1\over2}(2\pi)^2\Phi^{(n)}\Phi^{(-n)} \right]~~ } } The path
integral measure is modified by the FP determinant \ghostdetca. If
we diagonalize $X$ in terms of its eigenvalues $x_i$, the measure
factor involving $X$ becomes the measure for unitary matrices
\eqn\measx{ \prod_{i<j}\sin^2({x_i-x_j\over 2R}) } In other words,
the natural variable to be integrated over is the ``holonomy"
$U=e^{iX/R}$. As well known\refs{\Taylor,\CSmm}, this can be
thought of as the contribution from D-instantons at $x_i$ together
with all of their images at $x_i+2\pi nR$.

In the infinite radius limit $R\to\infty$, we can drop all the
modes $\Phi^{(n)}$ with nonzero winding since they become
infinitely ``massive". The action becomes simply \eqn\instinfca{ S
= {\beta\over 4\pi}{\rm Tr}([X,\Phi^{(0)}]^2 -
4\pi^2(\Phi^{(0)})^2). }

An alternative instanton matrix model for $c=1$ string, the
Liouville matrix model, was studied in \Mukhi. It would be nice to
understand its relation to the IMM we are proposing.

\subsec{The IMM for Type 0 String Theories}

Let us start with the MQM of type 0A theory in two dimensions,
which is the decoupled world volume theory of (stable) D0- and
anti-D0-branes. In the background with no net D0-brane charges,
the matrix model has $U(N)\times U(N)$ gauge symmetry. This is the
case we will be concerned with. We have the $U(N)\times U(N)$
gauge field $A_0$ and bifundamental tachyon $\phi$, \eqn\a{ A_0 =
\left( \eqalign{A ~~~~ 0 \cr 0 ~~~~ \tilde{A}} \right),~~~~ \phi =
\left( \eqalign{ 0 ~~~~ t \cr t^\dagger ~~~~ 0 } \right) } The
Lagrangian is \eqn\b{ L = {\rm Tr} \left[ (D_0 t)^\dagger (D_0 t)
+ {1\over 2\alpha'}t^\dagger t \right] } where $D_0t=\partial_0t +
At-t\tilde A$. Again, we want to compactify the Euclidean time on
a circle of radius $R'=\alpha'/R$, and rewrite the matrix quantum
mechanics in terms of a matrix integral over the Fourier modes of
$A$ and $t$. The D0- and anti-D0-branes becomes D- and
anti-D-instantons; the zero modes of $A$ and $\tilde A$ (Wilson
lines) become the collective coordinates of the D- and
anti-D-instantons; the Fourier modes of $t$ and $t^\dagger$ are
open strings stretched between D- and anti-D-instantons, which can
also wind around the circle. We expect this model to be equivalent
to type 0B MQM at radius $R$ in the double scaling limit.

Let us fix the gauge $\partial_0A_0=0$, which sets $A$ and $\tilde
A$ to their zero modes $A^{(0)}\equiv X/2\pi\alpha'$ and $\tilde
A^{(0)}\equiv \tilde X/2\pi\alpha'$. As before the gauge fixing
introduces the FP determinant \eqn\m{ \prod_{i<j}\left({\sin [
(x_i-x_j)/2R]\over (x_i-x_j)/2R}\right)^2 \left({\sin [ (\tilde
x_i-\tilde x_j)/2R]\over (\tilde x_i-\tilde x_j)/2R}\right)^2 }
where $x_i$ and $\tilde x_i$ are the eigenvalues of $X$ and
$\tilde X$ respectively. In terms of the Fourier modes $t^{(n)}$
of $t(x)$, the action is \eqn\n{ \eqalign{ S = {\beta\over
\pi(2\alpha')^{3/2}}\sum_{n\in{\bf Z}} {\rm Tr}\left[\left({2\pi n
R}t^{(n)}+Xt^{(n)}-t^{(n)}\tilde X\right) \left({2\pi n
R}t^{(n)\dagger}+t^{(n)\dagger}X-\tilde Xt^{(n)\dagger}\right)
\right.\cr\left. - a^2t^{(n)}t^{(n)\dagger}\right]~~ } } where we
exhibited the (type 0B) inverse ``Planck constant" $\beta$,
$a=\pi\sqrt{2\alpha'}$ is the critical distance. There was a
factor $2\pi \alpha'/R$ coming from the integral over Euclidean
time, but under T-duality $\beta\to \beta R/\sqrt{2\alpha'}$
\NewHat, so the factors of $R$ cancel out. If we further gauge fix
$X$ to the diagonal form, the usual measure factor
$\Delta(x)^2\Delta(\tilde x)^2$ is converted to the measure for
unitary matrices \eqn\o{ \prod_{i<j} \sin^2({x_i-x_j\over 2R})
\sin^2({\tilde x_i-\tilde x_j\over 2R}) } Therefore the natural
variables to be integrated over are the ``holonomies" $U=e^{2\pi
iX/R},~ \tilde U = e^{2\pi i\tilde X/R}$. The difference
disappears in the infinite radius limit $R\to \infty$. In this
limit all the winding modes are very massive and we can drop them,
so the action simplifies to \eqn\nactzt{ S = {\beta\over
\pi(2\alpha')^{3/2}} {\rm Tr}\left[(Xt-t\tilde X) (t^\dagger
X-\tilde Xt^{\dagger})
 - a^2tt^{\dagger}\right] }

Even though in the decompactification limit the integral over
$X,\tilde X$ is simply a Gaussian, we shouldn't integrate out them
directly. One reason is that, the determinant coming from
integration over $X,\tilde X$ would completely cancel the F-P
determinant coming from diagonalizing $t$. We would then naively
conclude that the eigenvalues of $t$ do not repel each other, and
it is unclear how to define the double scaling limit. What we have
done wrong is reminiscent to the case of gauged MQM, where
integrating out the gauge fields naively cancels the Vandermonde
determinants of the eigenvalues at every time. This is incorrect,
since if we carefully discretize the Euclidean time, the gauge
fields appear as ``link variables", while the tachyon fields are
associated with the ``sites". At the end, the F-P determinants of
the eigenvalues at the initial and end points of the time
evolution are not completely cancelled out. Their effect is
nothing but to antisymmetrize the wave functions. Similarly in the
IMM before integrating out $X$ and $\tilde X$, we should either
restrict to a finite interval of time, or compactify Euclidean
time on a finite circle.

There is another way to understand this subtlety. An unstable
D0-brane with $\lambda=1/2$ rolling tachyon profile is an array of
alternating D-instantons and anti-D-instantons along the thermal
circle, separated at the critical distance $a=\pi\sqrt{2\alpha'}$
\MSY. Suppose $2\pi R=2ma$ for some integer $m$. Naturally we
expect that the $U(N)$ MQM to be equivalent to the IMM with gauge
group $U(mN)\times U(mN)$. Therefore in the limit $R\to \infty$,
we must correspondingly take $N\to \infty$. We will come back to
this point in section 4.

Clearly we can interchange the role of type 0A and 0B MQM in the
above discussion. We will then obtain an IMM of type 0A theory.
Formally this model has the same action as the IMM for $c=1$
string. Presumably the double scaling limits are defined
differently in these two models.

\newsec{Integrable Structures in Two Dimensional String Theories}

It is well known that $c=1$ string deformed by closed string
momentum modes (or winding modes) have the integrable structure of
Toda chain hierarchy\refs{\DMR,\StringEq,\AKKtime,\KostovReview}.
The operators corresponding to the momentum modes generate Toda
flows, and the perturbed grand canonical partition function is
identified with a $\tau$-function. In this section we will show
that similar integrable structures appear in type 0A and 0B string
theories. Subsection 3.1 is a review of some results on the exact
S-matrix and integrable structures of ``theory I" ($c=1$ matrix
model), essentially following \refs{\AKKtime,\AKKnormal}. There is
nothing new in this first subsection, but it sets up the
conventions that will be used in the rest of the section. In
subsection 3.2, we study the S-matrix and integrable structures of
``theory II" (type 0B MQM). Most of these results are already
known, but we will derive the string equations for theory II
explicitly. In subsection 3.3, we show that type 0A MQM also has
the Toda integrable structure. This integrable structure is, in
some sense, more natural than that of ``theory I" since type 0A
theory is non-perturbatively well-defined. Subsection 3.4 studies
the perturbed partition function of type 0B IMM. We will indeed
recover the structure similar to that of type 0B MQM, and find a
dictionary translating closed string modes in spacetime to
operators in the IMM. In subsection 3.5, we compute the free
energy of IMM by explicitly performing the matrix integral. The
case for $c=1$ IMM was essentially done in
\refs{\UpsideDown,\KKK}. In the case of type 0B IMM, we find that
the free energy factorizes into two pieces, which precisely agree
with the contribution to type 0B MQM free energy from symmetric
and antisymmetric fluctuations of the fermi sea. By matching the
unperturbed free energies, we conclude that the string equations
for the integrable structure of IMM are the same as those for MQM.

\subsec{``Theory I"}

It is convenient to define light cone variables \eqn\lcvarcca{\hat
x_\pm = {\hat x\pm \hat p\over\sqrt{2}}, ~~~~~~[\hat x_+,\hat
x_-]=-i} In ``theory I" ($c=1$ string) these variables are
restricted to the region $x_\pm>0$, and for now we will not worry
about nonperturbative effects. The Hamiltonian of the free
fermions is \eqn\lchamca{ \hat H_0 = -{1\over 2}(\hat x_+\hat
x_-+\hat x_-\hat x_+) }  Since $\hat x_+,\hat x_-$ are conjugate
variables, we can write the wave function of a state in either
$x_+$ or $x_-$ representation. The two wave functions
$\psi_+(x_+)$ and $\psi_-(x_-)$ are related by a Fourier transform
\eqn\sopd{ \eqalign{ \psi_-(x_-)& = (\hat S \psi_+)(x_+) \cr &=
\int_0^\infty dx_+ K(x_-,x_+)\psi_+(x_+)} } where the integration
kernel $K$ is \eqn\defssad{ K(x_-,x_+)=\sqrt{2\over\pi}
\cos(x_-x_+) }  We could have also chosen
$K(z_-,z_+)=\sqrt{2/\pi}\sin(z_-z_+)$ instead. This ambiguity
reflects the fact that perturbatively the two sides of the fermi
sea decouple. The energy eigenstates have wave functions \eqn\wfa{
\psi^E_\pm(x_\pm) = {1\over \sqrt{2\pi}} x_\pm^{\pm iE-\half} }
Note that we have chosen a convenient basis for $\psi_+^E$ and
$\psi_-^E$, but they are {\sl not} related by \sopd. In fact, it
is straightforward to show that \eqn\sacts{ \hat
S^{\pm1}\psi_\pm^E(x_\pm) = {\cal R}^{\pm1}(E)
\psi_\mp^E(x_\mp),~~~~ {\cal R}(E) = \sqrt{2\over\pi}
\cosh({\pi\over 2}(i/2-E))\Gamma(iE+\half) } Essentially $\hat S$
is the S-matrix, and ${\cal R}(E)$ is the reflection coefficient
(phase shift).

The closed string momentum modes correspond to operators in MQM of
the form\refs{\Jevicki} \eqn\csmop{ V_{\pm k/R}(x) = e^{\pm
ikx/R}{\rm Tr} X_\pm(x)^{k/R} } where $x$ is the Euclidean time
variable, $R$ is the radius of the thermal circle, and the allowed
momenta are $\pm k/R$ for integer $k$. A general perturbation of
momentum modes is described by a potential \eqn\mompert{
V_\pm(x_\pm) = R\sum_{k\geq 1}t_{\pm k}x_\pm^{k/R} } In the $c=1$
MQM deformed by \mompert, the energy eigenstates are given by
``dressed" wave functions \eqn\dressa{ \Psi_\pm^E(x_\pm) = e^{\mp
i\varphi_\pm(x_\pm,E)}\psi_\pm^E(x_\pm) \equiv {\cal W}_\pm
\psi_\pm^E(x_\pm) } where the phases $\varphi_\pm$ are of the form
\eqn\phidefca{ \varphi_\pm(x_\pm;E) = V_\pm(x_\pm)+{1\over
2}\phi(E) - R\sum_{k\geq 1}{1\over k} v_{\pm k}(E)x_\pm^{-k/R} }
In above $\phi(E)$ is a constant phase shift. The terms involving
$v_{\pm k}$ vanish as $x_\pm \to \infty$. Semiclassically the
perturbation \mompert\ corresponds to deformed fermi sea profile
\eqn\dffs{\eqalign{ x_+x_- &= \mu + x_\pm
\partial \varphi_\pm(x_\pm;\mu) \cr &= \mu
+ \sum_{k\geq 1} kt_{\pm k} x_\pm^{k/R} + \sum_{k\geq 1} v_{\pm
k}(E)x_\pm^{-k/R} } } The compatibility of the two equations with
$+$ and $-$ signs in the subscripts puts constraints on $v_{\pm
k}$ in terms of $t_{\pm k}$'s. This comes from the requirement
that $\Psi_\pm^E$ are the wave functions of the same state in
$x_\pm$-representations. They are related by $\hat S \Psi_+^E =
\Psi_-^E$, or equivalently \eqn\keys{ {\cal W}_-={\cal W}_+
\hat{\cal R} } Note that this constraint is essentially determined
from the reflection coefficient ${\cal R}(E)$.

To see the integrable structure of Toda lattice hierarchy, we
shall recast the above in terms of operators on the $E$-space. For
example, it follows from \wfa\ that $\hat x_\pm$ are represented
as shift operators $\hat\omega^{\pm1}=e^{\mp i\partial_E}$. One
can define a Lax pair \eqn\laxpairca{ \eqalign{ &L_\pm = {\cal
W}_\pm \hat\omega^{\pm1} {\cal W}_\pm^{-1} = e^{\mp i\phi/2}
\hat\omega^{\pm1} \left( 1+\sum_{k\geq 1}a_{\pm k}\hat\omega^{\mp
k/R} \right) e^{\pm i\phi/2} \cr & M_\pm = -{\cal W}_\pm \hat E
{\cal W}_\pm^{-1} = \sum_{k\geq 1}kt_{\pm k}L_\pm^{k/R} -\hat E +
\sum_{k\geq 1}v_{\pm k}L_\pm^{-k/R} } } which satisfy the
commutation relation \eqn\laxcomm{ [L_\pm,M_\pm] = \pm iL_\pm }
Recall that the dressing operators ${\cal W}_\pm$ in terms of
$\hat E$ and $\hat \omega$ are of the form \eqn\dressopeo{ {\cal
W}_\pm = e^{\mp i\phi/2} \left( 1+\sum_{k\geq 1} w_{\pm
k}\hat\omega^{\mp k/R} \right) e^{\mp iR \sum_{k\geq 1} t_{\pm
k}\hat \omega^{\pm k/R}} } The integrable flow equation of the Lax
operators is \eqn\laxflow{
\partial_{t_n} L_\pm = [H_n,L_\pm] } where \eqn\laxhns{ H_n =
(\partial_{t_n}{\cal W}_+){\cal W}_+^{-1} = (\partial_{t_n} {\cal
W}_-){\cal W}_-^{-1} } By the virtue of \keys, the generator of
the Toda flow $H_n$ is the same for both $L_+$ and $L_-$. By a
standard argument (for example, see \AKKtime), it follows from the
structure of \dressopeo\ and \laxpairca\ that $H_n$ are of the
upper or lower triangular form \eqn\hndiag{ \eqalign{ &H_n =
(L_+^{n/R})_> + {1\over 2} (L_+^{n/R})_0, \cr &H_{-n} =
(L_-^{n/R})_< + {1\over 2} (L_-^{n/R})_0. }~~~~~~~~~~ (n>0) } It
also follows from \laxflow\ that the zero-curvature conditions on
$H_n$ are automatically satisfied.

The ``almost lower triangular" structure of lax operators $L_\pm$
\laxpairca, together with \laxflow\ and the important relation
\hndiag, define a Toda lattice hierarchy. The two Lax operators
$L_+[t_n]$ and $L_-[t_n]$ are in addition constrained by \keys.
Using the functional relation $R(E-i)=(-E+i/2)R(E)$, the
constraints on Lax operators can be expressed in terms of the so
called string equations, in this case given by \eqn\streqn{
\eqalign{ &M_+=M_-={1\over 2}(L_+L_-+L_-L_+), \cr &[L_+,L_-]=-i. }
} These conditions define a constrained Toda hierarchy.

The parameters $\phi,v_{\pm k}$ appearing in ${\cal W}_\pm$ are
determined in terms of $t_{\pm k}$ via the Toda flow \laxflow\ and
the initial condition $L_\pm|_{t_n=0}=e^{\mp
i\phi_0/2}\hat\omega^{\pm1} e^{\pm i\phi_0/2}$ where
$e^{i\phi_0(E)}={\cal R}(E)$. They are related to the Toda
$\tau$-function $\tau(E;t_n)$\foot{ The $\tau$-function appearing
here is denoted $\tau'[t]$ in \UenoTakasaki. In section 3.4 we
will introduce another $\tau$-function $\tau[t]$ defined using
vertex operators, following \refs{\CIS,\KKK}. They are related by
$\tau'[t]=\exp({\sum_{n>0}nt_nt_{-n}})\cdot\tau[t]$. The grand
canonical partition function is given by $\tau'[t]$. Also the
standard form of the $\tau$-function, $\tau_l[t]$, is related to
$\tau(E;t_n)$ by $\tau_l[t]=\tau(\mu+il/R;t_n)$. } by \eqn\taufn{
v_n = {\partial \ln\tau \over\partial t_n},~~~~ \phi(E) =
i\ln{\tau(E+ i/2R)\over\tau(E-i/2R)} } where we suppressed the
dependence on $t_n$'s. The $\tau$-function satisfies Hirota's
bilinear equations \refs{\Hirota,\UenoTakasaki}, an infinite set
of differential equations in $t_n$'s that completely determine
$\tau(E;t_n)$ provided the initial condition $\tau|_{t_n=0}(E)$
for all $E$. The initial condition for $\tau(E;t_n)$ is
essentially equivalent to the constraints imposed by the string
equations.

To see how the $\tau$-function is related to the free energy, let
us compute the density of states in terms of $\phi(E)$. We shall
introduce a cutoff at $x=\sqrt{2\Lambda}$, which is a wall that
reflects all momenta. In light cone variables this imposes a
boundary condition at $x_+=x_-=\sqrt\Lambda$, \eqn\bccool{
\Psi_+^E(\sqrt\Lambda)= \Psi_-^E(\sqrt\Lambda) } $\Psi_\pm^E$ as
defined in \dressa\ have asymptotic behavior \eqn\asymppsis{
\Psi_\pm^E(\sqrt{\Lambda}) \sim e^{\mp i \phi(E)/2}
(\sqrt{\Lambda})^{\pm iE}\times (E~{\rm independent~piece}) } It
follows form \bccool\ that the allowed energies $E_n$ satisfy
$i\phi(E_n)-iE_n\ln \Lambda=2\pi in$. The density of energy
eigenstates $\rho(E)$ is then \eqn\edens{ \rho(E) =
{\ln\Lambda\over 2\pi}-{1\over 2\pi }{d\phi(E)\over dE} } The free
energy is given by \eqn\freenrg{ \eqalign{ {\cal F} &=
\int_{-\infty}^\infty dE \rho(E) \ln (1+e^{-2\pi R(E+\mu)}) \cr &=
-R\int_{-\infty}^\infty dE {\phi(E)\over 1+ e^{2\pi R(E+\mu)}} \cr
&= i\sum_{n\geq 0} \phi\left({-\mu+(n+\half)i/ R}\right) \cr
&=\ln\tau(-\mu;t_n) } } where in the third line we closed the
contour in the upper half plane which picks up the poles at
$E=-\mu+(n+\half)i/R$, and in the last line we used the relation
\taufn. In our convention $\mu$ is the negative chemical
potential, which is positive when the fermi level is below the top
of the potential ($E=0$). We see that the perturbed grand
partition function is ${\cal Z}_\mu[t_n] = \tau(-\mu;t_n)$.

\subsec{``Theory II"}

Now let us turn to type 0B MQM, also known as ``theory of type
II"\MPR. For the unperturbed Hamiltonian there are two sets of
eigenfunctions that classically correspond to fermions in the left
and right sector of the fermi sea \eqn\obeigen{ \eqalign{
&\psi_{\pm,>}^E(x_\pm) = {1\over \sqrt{2\pi}} {x_\pm^{\pm
iE-\half}\over \sqrt{1+e^{2\pi E}}},~~~~(x_\pm>0) \cr
&\psi_{\pm,<}^E(x_\pm) = {1\over \sqrt{2\pi}} {(-x_\pm)^{\pm
iE-\half}\over \sqrt{1+e^{2\pi E}}},~~~~(x_\pm<0) } } In the other
half of the real $x_\pm$ axis the wave functions are defined by
analytic continuation, explicitly \eqn\anacon{ \eqalign{
&\psi_{\pm,>}^E(-x_\pm)=\pm ie^{\pi E}\psi_{\pm,>}^E(x_\pm),\cr
&\psi_{\pm,<}^E(x_\pm)=\pm ie^{\pi
E}\psi_{\pm,<}^E(-x_\pm),~~~~x_\pm>0. } } The wave functions in
$x_+$ and $x_-$ representations, $\psi_+$ and $\psi_-$, are not
merely related by a reflection because of the quantum tunnelling.
The operator $\hat S$ relating $\psi_+$ to $\psi_-$ is defined as
\sopd\ but with a different kernel \eqn\obker{ K(x_-,x_+)
={1\over\sqrt{2\pi}} e^{ix_+x_-} } $\hat S$ acts on the energy
eigenstates as \AKKtime \eqn\sactsb{ \left(\eqalign{\hat
S\psi^E_{+,>}(x_-)\cr \hat S \psi^E_{+,<}(x_-) }\right) = {\cal
R}(E) \left( \eqalign{ 1~~~~ & -ie^{\pi E} \cr -ie^{\pi E} &~~~~1
} \right)\left(\eqalign{\psi_{-,>}^E(x_-)\cr
\psi_{-,<}^E(x_-)}\right) } where \eqn\defra{ {\cal R}(E) =
{1\over \sqrt{2\pi}} e^{-{\pi\over 2}(E-i/2)}\Gamma(iE+\half) }
Perturbatively the left and right side of the fermi sea decouple.
This is reflected in the exponential suppression factor $e^{\pi
E}$ ($E<0$) when we analytically continue $x_\pm\to -x_\pm$. $\hat
S$ is diagonalized by symmetric and antisymmetric eigenfunctions
\eqn\sawaf{ \psi_{\pm,s}^E = {\psi_{\pm,>}^E+\psi_{\pm,<}^E\over
\sqrt{2}},~~~ \psi_{\pm,a}^E = {\psi_{\pm,>}^E-\psi_{\pm,<}^E\over
\sqrt{2}} } Then we have \eqn\sactb{\eqalign{ &\hat S
\psi_{+,s}^E(x_-) = R_s(E)\psi_{-,s}^E(x_-),~~~~ R_s(E) =
\sqrt{2\over \pi}\cosh[{\pi\over 2}(i/2-E)]\Gamma(iE+\half), \cr
&\hat S \psi_{+,a}^E(x_-) = R_a(E)\psi_{-,a}^E(x_-),~~~~ R_a(E) =
\sqrt{2\over \pi}\sinh[{\pi\over 2}(i/2-E)]\Gamma(iE+\half). } }
Note that $x_\pm$ act on $\psi_{\pm,>}^E$ as
$\hat\omega^{\pm1}=e^{\mp i\partial_E}$ and on $\psi_{\pm,<}^E$ as
$-\hat\omega^{\pm1}=-e^{\mp i\partial_E}$, or equivalently
\eqn\xactssa{ \eqalign{ & \hat x_\pm \psi_{\pm,s}^E =
\psi_{\pm,a}^{E\mp i},~~~~ \hat x_\pm \psi_{\pm,a}^E =
\psi_{\pm,s}^{E\mp i} } }

Now let us turn on perturbations by closed string momentum modes.
The dressed wave functions are of the form \eqn\dressedwaf{
\eqalign{ &\Psi_{\pm,s}^E(x_\pm) =
e^{\mp\varphi_{\pm}^s(\hat\omega^{\pm1};E)}\psi_{\pm,s}^E(x_\pm)\equiv
{\cal W}_\pm^s \psi_{\pm,s}^E(x_\pm), \cr &\Psi_{\pm,a}^E(x_\pm) =
e^{\mp\varphi_{\pm}^a(\hat\omega^{\pm1};E)}\psi_{\pm,a}^E(x_\pm)\equiv
{\cal W}_\pm^a \psi_{\pm,a}^E(x_\pm), } } where \eqn\phibs{
\eqalign{ &\varphi_\pm^s(\hat\omega^{\pm1};E) = R\sum_{k\geq
1}t_{\pm k}^s \hat\omega^{\pm k/R}+{1\over 2}\phi^s(E)
-R\sum_{k\geq 1} {1\over k}v_{\pm k}^s(E)\hat\omega^{\mp k/R} \cr
&\varphi_\pm^a(\hat\omega^{\pm1};E) = R\sum_{k\geq 1} t_{\pm k}^a
\hat\omega^{\pm k/R}+{1\over 2}\phi^a(E) -R\sum_{k\geq 1} {1\over
k} v^a_{\pm k}(E)\hat\omega^{\mp k/R} } } $t^s_{\pm k},t^a_{\pm
k}$ parameterize the symmetric and antisymmetric part of the
asymptotic perturbations of the fermi sea, respectively. ${\cal
W}_\pm$ again have the structure of \dressopeo. The Lax pairs
$(L_\pm,M_\pm)$ are defined the same way as \laxpairca, and
satisfy the relations \laxcomm, \laxflow, \laxhns. Now we have two
sets of generators of the flow, $H_n^s$ and $H_n^a$ (associated to
$t_n^a$ and $t_n^a$). Since $L_\pm$ and $\hat\omega$ are block
diagonal in the $(\psi_s,\psi_a)$ basis, \eqn\hnhh{ H_n=\left(
\eqalign{ H_n^s & ~~~0\cr ~0~~& ~~H_n^a} \right) } still satisfy
the upper or lower triangular relations \hndiag. They define two
independent Toda flows, associated to the symmetric and
antisymmetric perturbations of the fermi sea.

However, $\hat\omega^{\pm1}$ is no longer the same as $\hat x_\pm$
since the latter interchanges $\psi_s$ with $\psi_a$. Consequently
the functional constraint on the reflection coefficients is
modified to \eqn\fcrcob{ \eqalign{ R_s(E-i)=(-E+i/2)R_a(E), \cr
R_a(E-i)=(-E+i/2)R_s(E). } } So the strings equations \streqn\ no
longer hold in the case of ``theory II". They are modified to
\eqn\obstreqn{ \eqalign{ &M_+^{s,a}=M_-^{s,a},\cr &L_+^sL_-^s =
\tanh[{\pi\over2}({i/2}-M_\pm^s)]\, (-M^s_\pm+i/2),\cr &L_-^sL_+^s
= \coth[{\pi\over2}({i/2}-M_\pm^s)]\, (-M_\pm^s-i/2), \cr
&L_+^aL_-^a = \coth[{\pi\over2}({i/2}-M_\pm^a)]\,
(-M^a_\pm+i/2),\cr & L_-^aL_+^a =
\tanh[{\pi\over2}({i/2}-M_\pm^a)]\, (-M_\pm^a-i/2). } }
Perturbatively, i.e. in the limit $e^{\pi E}\ll 1$, these
equations reduce to \streqn. They define two independent
constrained Toda hierarchies, with the Lax operators acting on
$\psi_s$ and $\psi_a$ respectively. The perturbed grand canonical
partition function is the product of the two $\tau$-functions,
\eqn\tfn{ {\cal Z}_\mu[t^s,t^a]=\tau_s(\mu;t^s)\tau_a(\mu;t^a) }
As remarked in \DRSVW, this doesn't mean that the NS-NS and R-R
closed string modes decouple from each other. The symmetric and
antisymmetric perturbations of the fermi sea are mixtures of NS-NS
and R-R fields in spacetime, due to the nonlinearity of
bosonization.

\subsec{Type 0A theory}

Type 0A MQM can be represented by non-relativistic free fermions
moving in a two dimensional upside-down harmonic oscillator
potential. The Hamiltonian is \eqn\oaham{ \hat H = {1\over 2}(\hat
p_x^2+\hat p_y^2)-{1\over 4\alpha'} (\hat x^2+ \hat y^2) } The
theory has different independent sectors labelled by net D0-brane
charge $q$, which is the same as the angular momentum $\hat J =
\hat x \hat p_y-\hat y\hat p_x$ \NewHat. We shall mostly focus on
the case where there is no net D0-brane charge, namely the $J=0$
sector.

It is again convenient to define light cone variables \eqn\lcvar{
\hat x_\pm = {{1\over\sqrt{2\alpha'}}\hat x\pm \hat p_x\over \sqrt
2},~~~ \hat y_\pm = {{1\over\sqrt{2\alpha'}}\hat y\pm \hat
p_y\over \sqrt 2} } and \eqn\cplxlc{ \hat z_\pm = \hat x_\pm +i
\hat y_\pm,~~~ \hat{\bar z}_\pm = \hat x_\pm - i\hat y_\pm } We
have commutators \eqn\comms{ \eqalign{ &[\hat x_+,\hat x_-]=[\hat
y_+,\hat y_-]=i/\sqrt{2\alpha'},\cr &[\hat z_+,\hat
z_-]=[\hat{\bar z}_+, \hat{\bar z}_-]=0, \cr &[\hat z_+,\hat {\bar
z}_-]=[\hat{\bar z}_+, \hat z_-]=2i/\sqrt{2\alpha'}. }} In light
cone variables the Hamiltonian is written as \eqn\hamlc{ \eqalign{
\hat H &= -(\hat x_+\hat x_- +\hat y_+\hat
y_--{i\over\sqrt{2\alpha'}}) \cr &= -{1\over 2}(\hat z_+\hat {\bar
z}_- + \hat{\bar z}_+ \hat z_- -{2i\over\sqrt{2\alpha'}}) \cr &=
{i\over \sqrt{2\alpha'}} (z_+{\partial\over \partial z_+} + \bar
z_+ {\partial\over
\partial\bar z_+}+1) } } where the last line is written in
$(z_+,\bar z_+)$-representation. In addition, we have commutation
relations \eqn\angularcomm{ [\hat J,\hat z_\pm]=z_\pm,~~~ [\hat
J,\hat{\bar z}_\pm]=-\hat{\bar z}_\pm, } or in $(z_+,\bar z_+)$
representation \eqn\jzz{ \hat J = z_+{\partial\over \partial z_+}
-\bar z_+ {\partial\over \partial \bar z_+} } The wave function of
a state can be expressed either in $(z_+,\bar z_+)$ representation
or in $(z_-,\bar z_-)$ representation, denoted by $\psi_+(z_+,\bar
z_+)$ and $\psi_-(z_-,\bar z_-)$ respectively. Since we restrict
our wave functions to have zero angular momentum, it must be of
the form $\psi_\pm(z_\pm,\bar z_\pm)=\psi_\pm(z_\pm \bar z_\pm)$.
The energy eigenstates are given by (setting $\alpha'=2$)
\eqn\eigennrgst{ \psi_\pm^E \sim z_\pm^{\pm iE-\half} {\bar
z}_\pm^{\pm iE-\half} } The wave functions in $(z_+,\bar z_+)$ and
$(z_-,\bar z_-)$ representations are related by \eqn\szz{
\eqalign{ \psi_-(z_-,\bar z_-) &= (\hat S \psi_+)(z_-,\bar z_-)
\cr &= \int dz_+ d\bar z_+ K(\bar z_-,z_+) K(z_-,\bar z_+)
\psi_+(z_+,\bar z_+) } } where $K(z_-, z_+) = {1\over \sqrt{2\pi}}
e^{iz_- z_+}$. Acting on energy eigenstates, we have
\eqn\actnrgess{ \hat S \psi_+^E = {\cal R}(E) \psi_-^E,~~~~ {\cal
R}(E) = {\Gamma(iE+\half)\over \Gamma(-iE+\half)} } This is the
same as the phase shift found in \NewHat.

Now we can study perturbations of the fermi sea. Let us consider
dressed wave functions \eqn\oadressed{ \Psi_\pm^E = e^{\mp
\varphi(z_\pm \bar z_\pm;E)} \psi_\pm^E \equiv {\cal W}_\pm
\psi_\pm^E } where the phases $\varphi_\pm$ have Laurent expansion
\eqn\oaphase{ \varphi_\pm(z_\pm\bar z_\pm;E) = {1\over 2}\phi(E) +
R\sum_{k\geq 1} t_{\pm k} (z_\pm\bar z_\pm)^{k/R} -R \sum_{k\geq
1} {1\over k} v_{\pm k} (z_\pm \bar z_\pm)^{-k/R} } $t_{\pm k}$
parameterize the asymptotic perturbation by momentum modes of
NS-NS scalars, corresponding to the operator \eqn\momoav{
V_{p=k/R}={\rm Tr} (Z_\pm \bar Z_\pm)^{|k|/R}} in type 0A MQM,
where the sign of the subscripts depends on the sign of $k$.
$(\hat z_\pm \hat{\bar z}_\pm)$ can be represented as shift
operators $\hat\omega^{\pm 1}$, where $\hat\omega$ acts on energy
eigenstates as $\hat \omega^{\pm1}\psi_\pm^E = \psi_\pm^{E\mp i}$.
We have $\hat\omega=e^{-i\partial_E}$ and the commutation
relations \eqn\shiftoa{ [\hat \omega^{\pm}, -\hat E] =\pm
i\hat\omega^{\pm} } As before we can define a Lax pair \eqn\oalax{
L_\pm = {\cal W}_\pm \hat\omega^\pm {\cal W}_\pm^{-1},~~~ M_\pm =
-{\cal W}_\pm \hat E {\cal W}^{-1}_\pm } The dressing operators
${\cal W}_\pm$ satisfy the constraint \eqn\wsoa{ {\cal W}_-= {\cal
W}_+ \cdot{\cal R}(\hat E) } This again defines the structure of
constrained Toda lattice hierarchy. But the string equation
\streqn\ is modified to the following \eqn\oastreqn{ \eqalign{
M_+=M_-,~~~ [L_+,L_-]= 2iM_\pm,~~~\{L_+,L_-\}=2M_\pm^2-{1\over2}.
} }


The density of states, and hence the free energy, is related to
the phase $\phi(E)$ in the standard way. To compute the density of
states, we shall introduce a cutoff at $x^2+y^2=\Lambda$. The
cutoff wall reflects all the momenta, so we have $xp_x+yp_y=0$ as
well. Further we demand the vanishing of angular momentum
$xp_y-yp_x=0$. The combination of these is equivalent to $z_+\bar
z_+=z_- \bar z_-=\Lambda$. We can impose a boundary condition at
the wall \eqn\consoa{ \Psi_+^E(\Lambda) = \Psi_-^E(\Lambda) } It
follows that the density of states depends on $E$ the same way as
in \edens.

It is not hard to generalize the above construction to sectors of
nonzero net D0-brane charge $q$. These backgrounds are identified
as extremal black holes in type 0A string theory\BHOA. In this
case, the energy eigenstates (carrying angular momentum $q$) are
\eqn\eigmq{ \psi_\pm^E \sim z_\pm^{\pm iE+{q/ 2}-\half} \bar
z_\pm^{\pm iE-{q/ 2}-\half} } The reflection coefficients are
computed from \szz\ to be \eqn\rcmq{ {\cal R}(E) =
{\Gamma(iE+{q+1\over 2})\over\Gamma(-iE+{q+1\over2})} } The Lax
operators are defined as before, but the string equations become
\eqn\newsemq{ M_+=M_-,~~~[L_+,L_-]=2iM_\pm,~~~\{L_+,L_-\}=2M_\pm^2
+{q^2-1\over 2} . } Interestingly, $\{L_+,L_-,M\equiv M_\pm\}$
form a representation of $sl(2,{\bf R})$, with isospin
$l={(|q|-1)/2}$.


\subsec{Type 0B IMM perturbed by closed string momentum modes}

Following \refs{\UpsideDown,\KKK}, we shall integrate out the
tachyons $t^{(n)}$, $t^{(n)\dagger}$ in the type 0B IMM \n\ by
analytic continuation, and obtain a matrix integral in terms of
the eigenvalues of $X$ and $\tilde{X}$, \eqn\matint{ \eqalign{
&\int \prod dx_i d\tilde x_i \prod_{i<j} \sin^2 ({x_i-x_j\over
2R}) \sin^2 ({\tilde x_i-\tilde x_j\over 2R})
\prod_{i,j}\prod_{n=-\infty}^{\infty}{1\over (2\pi nR + x_i
-\tilde x_j)^2 - a^2} \cr &= \int \prod dx_i d\tilde x_i
\prod_{i<j}\sin^2 ({x_i-x_j\over 2R}) \sin^2 ({\tilde x_i-\tilde
x_j\over 2R})\prod_{i,j}{1\over \sin^2 [( x_i-\tilde x_j)/ 2R]-
\sin^2 (a/2R) } } } Roughly speaking we have a system of $N$
eigenvalues $x_i$ and $\tilde{x}_j$, they repel eigenvalues of the
same type through the F-P determinant and interact through some
effective potential. The analytic continuation made above is quite
naive. As a trade off, \matint\ is not unambiguously defined, and
we have to give a correct contour prescription. This will be
considered in the next section, for now we still formally work
with \matint.

Let us write the partition function in terms of the integral over
eigenvalues of $U,\tilde U$, $z_j=e^{ix_j/R}, \tilde
z_j=e^{i\tilde x_j/R}$, \eqn\tparta{ \eqalign{ Z_N &= {1\over
(N!)^2} \prod_{k=1}^N \oint {dz_k\over 2\pi i z_k} \oint {d\tilde
z_k\over 2\pi i \tilde z_k} \prod_{i<j}|z_i-z_j|^2 |\tilde
z_i-\tilde z_j|^2 \cr &~~~~~~\times \prod_{i,j}{1\over |z_i
q^{1/2}-\tilde z_j q^{-1/2}|\cdot |z_i q^{-1/2}-\tilde z_j
q^{1/2}|}\cr &= {1\over (N!)^2} \prod_{k=1}^N \oint {dz_k\over
2\pi i} \oint {d\tilde z_k\over 2\pi i} \prod_{i\not=j}{(z_i-z_j)
(\tilde z_i-\tilde z_j)}\cr &~~~~~~\times \prod_{i,j}{1\over (z_i
q^{1/2}-\tilde z_j q^{-1/2}) (z_i q^{-1/2}-\tilde z_j q^{1/2})}. }
} where $q=e^{ia/R}=e^{\pi i\sqrt{2\alpha'}/R}$. Consider the
perturbation by momentum modes (winding modes in the T-dual type
0A theory) \eqn\windpar{ \sum_{n\in {\bf Z}}\lambda_n {\rm Tr} U^n
+\tilde\lambda_n {\rm Tr}\tilde U^n } The generating functional is
then \eqn\genfn{ \eqalign{ Z_N[\lambda,\tilde \lambda] &= {1\over
(N!)^2} \prod_{k=1}^N \oint {dz_k\over 2\pi i} \oint {d\tilde
z_k\over 2\pi i} e^{u(z_k)+\tilde u(\tilde
z_k)}\prod_{i\not=j}{(z_i-z_j) (\tilde z_i-\tilde z_j)}\cr
&~~~~~~\times \prod_{i,j}{1\over (z_i q^{1/2}-\tilde z_j q^{-1/2})
(z_i q^{-1/2}-\tilde z_j q^{1/2})} \cr &= {1\over (N!)^2}
\prod_{k=1}^N \oint {dz_k\over 2\pi i} \oint {d\tilde z_k\over
2\pi i} e^{u(z_k)+\tilde u(\tilde z_k)} \cr &~~~~~~\times
{\det}_{ij} \left({1\over z_i q^{1/2}-\tilde z_j q^{-1/2}}\right)
 {\det}_{ij} \left({1\over z_i q^{-1/2}-\tilde z_j
q^{1/2}}\right) } } where \eqn\uut{ u(z) = \sum_n\lambda_n z^n,~~~
\tilde u(\tilde z) = \sum_n\tilde \lambda_n \tilde z^n, } and we
have used the Cauchy identity \eqn\cauchyid{
{\Delta(a)\Delta(b)\over \prod_{i,j}(a_i-b_j)} = {\det}_{ij}\left(
{1\over a_i-b_j} \right) } It is most convenient to consider the
grand canonical partition function \eqn\grandc{
Z_\mu[\lambda,\tilde \lambda] =\sum_{N=0}^\infty
e^{\pi\sqrt{2\alpha'} \mu N} Z_N[\lambda,\tilde \lambda] } where
$\mu$ is the chemical potential of type 0B theory, related to the
one of the T-dual type 0A theory by $\mu'=\mu R/\sqrt{2\alpha'}$.
For reasons that will become clear shortly, let us define
\eqn\tuu{ \lambda_n=2t_n+(q^n+q^{-n})\tilde t_n,~~~ \tilde
\lambda_n= -2\tilde t_n -(q^n+q^{-n})t_n } and a
``$\tau$-function" \eqn\taufn{ \eqalign{ \tau_l[t,\tilde t] &=
e^{-\sum_n n[2t_n t_{-n}+2\tilde t_n \tilde
t_{-n}+(q^n+q^{-n})(t_n\tilde t_{-n}+\tilde t_{n} t_{-n})]}
\sum_{N=0}^\infty (q^{2l} e^{\pi\sqrt{2\alpha'}\mu})^N
Z_N[t,\tilde t]\cr &= e^{-\sum_n n[2t_n t_{-n}+2\tilde t_n \tilde
t_{-n}+(q^n+q^{-n})(t_n\tilde t_{-n}+\tilde t_n t_{-n})]}
Z_{\mu+2il/R}[t,\tilde t] } } One might hope that the grand
canonical partition function $Z_\mu[t,\tilde t]$ is the
$\tau$-function of some integrable hierarchy. This is not quite
true. The perturbations that generate the integrable flows are
related to $t_n,\tilde t_n$'s through some bosonization maps, as
we will show below.

It is useful to rewrite the partition function in vertex operator
formalism, analogous to the case of $c=1$ string \KKK. One can
introduce two independent 2D chiral bosons $\varphi_{1,2}(z)$,
with mode expansion \eqn\modeexp{\varphi_{1,2}(z) = \hat
q_{1,2}+\hat p_{1,2} \ln z + \sum_{n\not=0} {H_n^{(1,2)}\over
n}z^{-n} } The vacuum $|l\rangle$ is defined by \eqn\vacl{
H_n^{(1),(2)}|l\rangle=0~~~(n>0),~~~~ \hat p_1|l\rangle = \hat
p_2|l\rangle=l|l\rangle. } We could have considered more general
vacuum state $|l_1,l_2\rangle$, but that would be unnecessary for
our purpose. We further define \eqn\newphis{ \eqalign{ &\phi(z) =
\varphi_1(q^{1/2}z)-\varphi_2(q^{-1/2}z),\cr &\tilde \phi(z) =
\varphi_1(q^{-1/2}z)-\varphi_2(q^{1/2}z), } } and $H_n,\tilde H_n$
the corresponding creation and annihilation operators in the mode
expansion of $\phi,\tilde\phi$, \eqn\htildeh{ \eqalign{ & H_n=
q^{-n/2} H_n^{(1)} -q^{n/2} H_n^{(2)} \cr & \tilde H_n= q^{n/2}
H_n^{(1)} -q^{-n/2} H_n^{(2)} } } Note that $\phi$ and
$\tilde\phi$ are not independent fields. We have commutation
relations \eqn\commhns{ [H_n,H_m] = 2n\delta_{n+m}=[\tilde
H_n,\tilde H_m],~~~~~ [H_n,\tilde H_m] =
(q^n+q^{-n})n\delta_{n+m}. } Using the operator \eqn\opg{ \hat{\bf
g}' = \sum_{N=0}^\infty {1\over (N!)^2}\left(q^{-i\mu R}\oint
{dz\over2\pi}:e^{\phi(z)}: \oint{d\tilde z\over
2\pi}:e^{-\tilde\phi(\tilde z)}:\right)^N } the ``$\tau$-function"
\taufn\ can be written as \eqn\taucpt{ \tau_l[t,\tilde t] =
\langle l| e^{-\sum_{n>0} {t_nH_n+\tilde t_n \tilde H_n} }
\hat{\bf g}'\, e^{\sum_{n>0} {t_{-n}H_{-n}+\tilde t_{-n} \tilde
H_{-n}}}|l\rangle } To see this, observe that the contractions
among the operators $e^\phi$ and $e^{-\tilde\phi}$ in \opg\ give
the integrand in \tparta, and commuting ${\bf g}'$ through
$e^{-\sum_{n>0} t_nH_n+\tilde t_n \tilde H_n}$ and $e^{\sum_{n<0}
t_nH_n+\tilde t_n \tilde H_n}$ give the momentum mode perturbation
$e^{u+\tilde u}$ appearing in \genfn. The symmetry
$\phi\to\phi+a$, $\tilde\phi\to\tilde\phi+a$ is respected by the
vacuum state $|l\rangle$. Therefore, we can replace $\hat {\bf
g}'$ by \eqn\gnewall{ \hat {\bf g} = \exp\left(q^{-i\mu R/2}\oint
{dz\over2\pi}:e^{\phi(z)}+e^{-\tilde\phi(z)}:\right) } in the
partition function \taucpt. The partition function in sectors of
nonzero net D-instanton number can be obtained from the general
vacuum state $|l_1,l_2\rangle$. The latter are closely related to
the solitonic sectors of the type 0B bosonic Hilbert space
discussed in \DRSVW. Of course, the $\tau$-function \taucpt\ does
not simply factorize into two components involving only $t_n$ and
$\tilde t_n$ respectively, because $\phi$ and $\tilde\phi$ have
nontrivial OPE.

The exponent in $\hat{\bf g}$ can be interpreted as a Hamiltonian.
It is clearer to fermionize $\varphi_1,\varphi_2$: \eqn\bosonphis{
\eqalign{ & e^{\varphi_1(z)}\simeq \psi_1(z),~~~
e^{-\varphi_1(z)}\simeq \bar\psi_1(z), \cr &
e^{\varphi_2(z)}\simeq \psi_2(z),~~~ e^{-\varphi_2(z)}\simeq
\bar\psi_2(z). } } The vacuum $|l\rangle$ satisfies \eqn\vacfermi{
\psi_{1,r}|l\rangle = \psi_{1,-r}^\dagger|l\rangle =
\psi_{2,r}|l\rangle = \psi_{2,-r}^\dagger|l\rangle=0,~~~r\in{\bf
Z}+\half,~r>l. } In terms of $\psi_{1,2}$, we can write
\eqn\gbose{ \hat{\bf g} = \exp\left\{ q^{-i\mu R/2} \oint {dz\over
2\pi}\left[- \bar\psi_2(q^{-1/2}z)\psi_1(q^{1/2}z) +
\bar\psi_1(q^{-1/2}z)\psi_2(q^{1/2}z) \right] \right\} } We can
interpret \taucpt\ as a partition function of the fermions
$\psi_1,\psi_2$. The integral in the exponent of $\hat{\bf g}$ can
be written in first-quantized form as \eqn\firstham{\eqalign{\hat
H &= e^{{i\over 2}\ln q(\hat z\hat p_z+\hat p_z\hat z)}P_{12}
\cr&= e^{-{a\over 2R}(\hat z\hat p_z+\hat p_z \hat z)}P_{12} } }
where $P_{12}: \psi_1\to\psi_2,\psi_2\to-\psi_1$, and we have used
$q=e^{ia/R}$. This ``Hamiltonian" can be compared to the $c=1$
string case\KKK, where one can write the partition function in
terms of vertex operators involving a single chiral boson
$\varphi(z)$ or its fermionization $\psi(z)$, and the
corresponding ``Hamiltonian" is $\hat H = e^{-{a\over 2R}(\hat
z\hat p_z+\hat p_z \hat z)}$.

Since $\psi_\pm={1\over\sqrt{2}}(\psi_1\pm i\psi_2)$ diagonalize
$P_{12}$, it is clear that $\tau_l[t_n=\tilde t_n=0]$ factorizes
as the product of partition functions involving $\psi_+$ and
$\psi_-$ separately. In the type 0B fermi sea picture,
$\psi_+,\psi_-$ should correspond to symmetric and antisymmetric
perturbations of the fermi sea. Indeed, as we will show in the
next section, the partition function involving $\psi_+$ ($\psi_-$)
agrees with the partition functions in type 0B MQM involving only
symmetric (antisymmetric) perturbations.

In the Hamiltonian interpretation, the ``initial" state in the
expression \taucpt\ for the $\tau$-function is \eqn\instate{
e^{\sum_{n>0}{t_{-n}H_{-n}+\tilde t_{-n}\tilde H_{-n}} }|l\rangle
=
e^{\sum_{n>0}t_{-n}^{(1)}H_{-n}^{(1)}+t_{-n}^{(2)}H_{-n}^{(2)}}|l\rangle
} where \eqn\sit{t_n^{(1)} = {q^{-n/2}t_n+q^{n/2}\tilde
t_n},~~~t_n^{(2)} = {-q^{-n/2}\tilde t_n-q^{n/2} t_n}} As we have
seen, $\psi_1\pm i\psi_2$ correspond to symmetric and
antisymmetric perturbations of the fermi sea. It is natural to
expect that $\psi_1$ and $\psi_2$ correspond to perturbations of
the left and right sector of the fermi sea (decoupled
perturbatively). Consequently the bosonized modes
$H_n^{(1)},H_n^{(2)}$ are linear combinations of momentum modes of
NS-NS and R-R scalars in spacetime. From \tuu\ and \sit, the
perturbation parameters $t_n^{(1)}, t_n^{(2)}$ are related to
$\lambda_n,\tilde\lambda_n$ in IMM by \eqn\ttll{ \eqalign{
&\lambda_n = q^{n/2} t_n^{(1)} - q^{-n/2} t_n^{(2)} \cr & \tilde
\lambda_n = -q^{-n/2} t_n^{(1)} + q^{n/2} t_n^{(2)} } } This also
leads us to the identification \eqn\idsa{ \eqalign{ & a_{L,n}\sim
q^{n/2}{\rm Tr} e^{inX/R} - q^{-n/2}{\rm Tr} e^{in\tilde X/R}, \cr
& a_{R,n}\sim -q^{-n/2}{\rm Tr} e^{inX/R} + q^{n/2}{\rm Tr}
e^{in\tilde X/R}, } } or \eqn\idsaor{ \eqalign{ & {\rm
Tr}e^{inX/R}\sim {q^{n/2}a_{L,n} + q^{-n/2}a_{R,n}\over
q^n-q^{-n}}, \cr & {\rm Tr}e^{in\tilde X/R}\sim {q^{-n/2}a_{L,n} +
q^{n/2}a_{R,n}\over q^n-q^{-n}}, } }where $a_{L,n}\pm a_{R,n}$ are
the momentum modes of the NS-NS and R-R scalar with momenta
$p=n/R$. Note that the Lorentzian energy is $E=in/R$. To
analytically continue to Lorentzian signature, we have
$q^n=e^{2\pi E}$. Perturbatively positive powers of $q$ can be
neglected ($E<0$ below the top of the potential). In this limit we
have approximately ${\rm Tr}e^{inX/R}\sim -e^{\pi E}a_{R,n},~{\rm
Tr}e^{in\tilde X/R}\sim -e^{\pi E}a_{L,n}$. As will be shown in
section 5, the relations \idsa\ precisely reproduce the ZZ
boundary state in super-Liouville theory.

We should remind the reader that $H^{(1)}_n\pm iH_n^{(2)}$ are
{\sl not} the bosonization of $\psi_1\pm i\psi_2$, and that
$t_n^{(1)}\pm it_n^{(2)}$ are not the same as $t^s_n,t^a_n$
defined in section 3.2. The operators ${\rm Tr}e^{inX/R},{\rm
Tr}e^{in\tilde X/R}$ in IMM correspond to linear combinations of
NS-NS and R-R modes, whereas in MQM the fluctuations of the fermi
sea are related to closed string fields in spacetime by
bosonization. Due to this bosonization relation between
$t_n,\tilde t_n$ and $t_n^s,t_n^a$, the partition function of IMM
does not factorize in a manifest way. However if we replace the
``initial state" (and similarly the ``final state") in \taucpt\ by
\eqn\bosh{ e^{\sum_{n>0} t_{-n}^s H_{-n}^++ t_{-n}^a H_{-n}^-}
|l\rangle } where $H_n^{\pm}$ are the modes of the bosonization of
$\psi_\pm$, then the partition function will factorize into two
$\tau$-functions of constrained Toda hierarchies that depend only
on $t^s$ or $t^a$, just as in type 0B MQM. To show that the
perturbed grand partition functions in type 0B IMM and MQM are
actually the same, it remains to show that the constraints (string
equations) of the IMM agree with those of MQM.

\subsec{String equations for IMM}

In this subsection we derive the string equations for type 0A and
0B IMM, by computing the commutator $[L_+,L_-]$ at $t_n=0$. Using
the general form of Lax operators \laxpairca, we know that in the
absence of perturbation, \eqn\lo{ \eqalign{ &L_\pm = e^{\mp
i\phi_0/2} \hat\omega^{\pm1} e^{\pm i\phi_0/2}, \cr &[L_+,L_-] =
e^{i\phi_0(E+i)-i\phi_0(E)} - e^{i\phi_0(E)-i\phi_0(E-i)}. } }
Using the relation \freenrg, we have \eqn\phif{ \phi(\mu) =
-i\left[{\cal F}(\mu+{i\over 2R}) - {\cal F}(\mu-{i\over
2R})\right] } where ${\cal F}(\mu)=\ln {\cal Z}(\mu)$ is the
unperturbed free energy. To show that the IMM string equations are
the same as the ones for MQM found in previous sections, it
suffices to show that $\phi(\mu)$ as given by \phif\ (with ${\cal
F}$ evaluated from IMM) does lead to the correct reflection
coefficient ${\cal R}(E)=e^{i\phi(E)}$.

Let us first consider $c=1$ or type 0A IMM, which is already done
in \UpsideDown. The grand partition function takes the form of a
Freholm determinant \eqn\cagp{ {\cal Z}(\mu) = \det (1+q^{i\mu R
}\hat K) } where $q=e^{2\pi i/R}$, $\mu$ is the chemical potential
of the theory on radius $R$ (the chemical potential of the T-dual
theory is $\mu'=\mu R$). $\hat K$ is defined by \eqn\fredker{
(\hat K f)(z) = -\oint {dz'\over 2\pi i} {f(z')\over
q^{1/2}z-q^{-1/2}z'} } The contour prescription of \UpsideDown\
(see also \KostovReview) is to add a small imaginery part to $R$
so that $|q|<1$. A basis that diagonalizes $\hat K$ consists of
the monomials $z^n$, with \eqn\Kzn{ \hat K z^n = \left\{ \eqalign{
& q^{n+\half}z^n,~~~n\geq 0 \cr & 0,~~~~~~~n<0 } \right. } The
free energy is given by \eqn\zfscaimm{ \eqalign{ {\cal F}(\mu) &=
\sum_{n\geq 0} \ln(1+q^{i\mu R+n+\half}) \cr &= \sum_{m=1}^\infty
{(-1)^m\over m} {q^{im\mu R}\over q^{m/2}-q^{-m/2}} \cr &=\int_C
{dt\over t} {e^{i\mu t}\over 4\sinh (t/2) \sinh (t/2R)} } } where
$C$ is the contour that picks up the poles of $\sinh(t/2)$ at
$t=2\pi ni$ ($n>0$). For example, we can choose $C$ to run from
${-\infty}$ to $0$ along the real axis, and from $0$ to $i\infty$
on the right of the imaginery axis, as shown in Fig. 2. \fig{The
integration contour $C$ in the complex $t$-plane that appears in
\zfscaimm, where $R$ is given a small negative imaginery
part.}{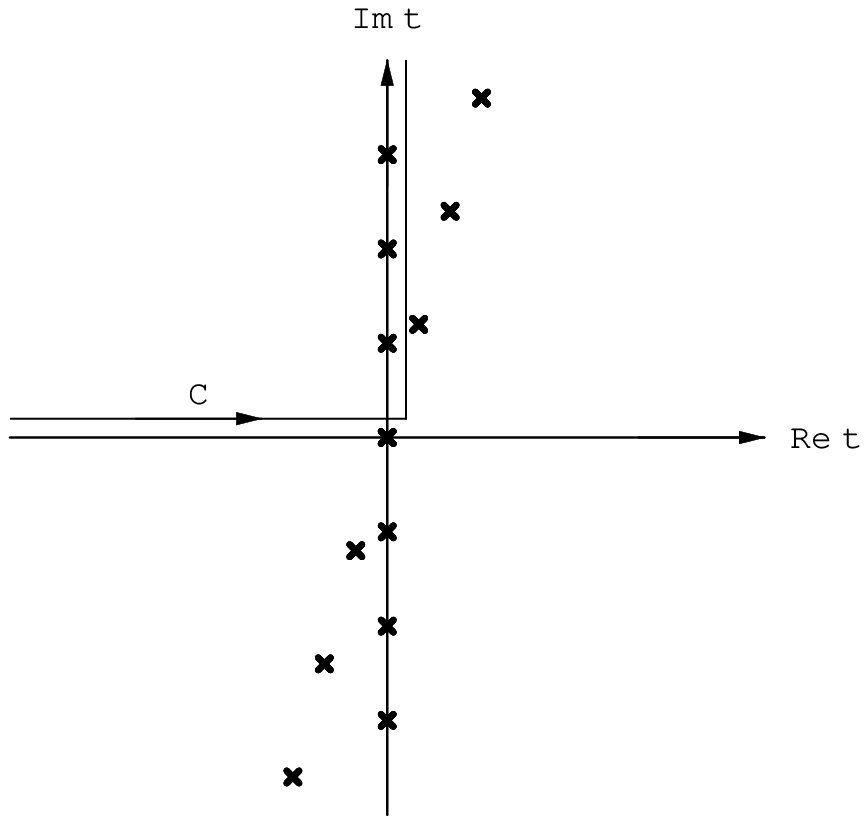}{2.5truein} \zfscaimm\ is however not quite
right. It is in fact the free energy of a system with ordinary
(``right sign") harmonic oscillator potential. Nevertheless, the
integrand in the last line of \zfscaimm\ has the same structure as
in the free energy computed from the true spectrum of the
upside-down harmonic oscillator potential. This suggests that we
modify the contour prescription, so that \eqn\crfn{ {\cal F}={\rm
Re}\int_0^\infty {dt\over t} {e^{i\mu t}\over 2\sinh(t/2)
\sinh(t/2R)} } It then follows from \phif\ that \eqn\phisfz{
\eqalign{ \phi(\mu)&= -\int_0^\infty {dt\over t} {\sin(\mu t)\over
\sinh (t/2)  } \cr &= -i \ln {\Gamma(i\mu+\half)\over
\Gamma(-i\mu+\half)} } } which agrees with the type 0A MQM
reflection coefficient \actnrgess.

The change of contour in above calculation, which effectively
takes us from the right sign harmonic oscillator potential to the
wrong sign potential, seems rather ad hoc. It will be nice to have
a clearer prescription. We will however use the same prescription
to compute the free energy for type 0B IMM, and find a nontrivial
agreement with the results obtained from the MQM.

In the case of type 0B IMM, we have seen from the previous
subsection that the unperturbed partition function is the product
of two Fredholm determinants, \eqn\cagpob{ {\cal Z}(\mu) =
\det(1+iq^{i\mu R/2} K) \det(1-iq^{i\mu R/2}K) } where again,
$\mu'=\mu R$ is the chemical potential of the T-dual type 0A MQM.
The ``symmetric part" and ``antisymmetric part" of the free energy
are \eqn\safnrg{ \eqalign{ {\cal F}_s(\mu) &= \sum_{n\geq 0}\ln
(1+iq^{i\mu R/2 + n +\half}) \cr &=\int_C {dt\over t} {e^{i\mu
t/2} e^{t/4} \over 4\sinh(t/2)\sinh(t/2R)}, \cr {\cal F}_a(\mu) &=
\sum_{n\geq 0}\ln (1-iq^{i\mu R/2 + n +\half}) \cr &= \int_C
{dt\over t} {e^{i\mu t/2} e^{3t/4} \over 4\sinh(t/2)\sinh(t/2R)}.
} } where $C$ is the contour that picks up the poles of
$\sinh(t/2)$ at $2\pi ni$ ($n>0$). Again, with proper modification
of the contour prescription, and using \phif, we obtain the phase
shifts \eqn\phan{ \eqalign{ \phi_s(\mu) &= -\int_0^\infty {dt\over
t} {\sin(\mu t/2) e^{t/4}\over \sinh(t/2)} \cr &= -i\ln
{\Gamma({i\mu\over2}+{1\over4})\over
\Gamma(-{i\mu\over2}+{1\over4})}, \cr \phi_a(\mu) &=
-\int_0^\infty {dt\over t} {\sin(\mu t/2) e^{3t/4}\over
\sinh(t/2)} \cr &= -i\ln {\Gamma({i\mu\over2}+{3\over4})\over
\Gamma(-{i\mu\over2}+{3\over4})}. } } Up to terms that contribute
a constant to the density of states, these precisely reproduce the
reflection coefficients \sactb.

Note that one can assign different chemical potentials $\mu_\pm$
to the symmetric and antisymmetric perturbations. This corresponds
to turning on a background RR flux \Gross. The T-dual type 0A flux
background should be analogously defined by giving different
chemical potentials to two independent winding sectors. In this
case we need to work with the grand canonical ensemble which sums
up sectors of arbitrary D0-brane and anti-D0-brane numbers. We
don't expect the corresponding type 0A background to have a fixed
net D0-brane charge $q$. The thermal fluctuation in $q$ should be
of order ${\cal O}(e^{-\pi\beta\mu})$ instead of ${\cal
O}(e^{-2\pi\beta\mu})$, since the ``effective" chemical potential
for the two winding sectors in type 0A theory on radius $1/R$ is
$\mu R/2$ ($\mu$ being the chemical potential in the T-dual type
0B theory), as in  \cagpob. This is responsible for the mismatch
at nonperturbative level between type 0A partition function with
$\mu_\pm=\mu\pm Q$ and type 0B partition function with RR flux
$q=iQ$ as found in \Gross.

With proper contour prescription for the type 0A and 0B IMM, we
have produced the correct reflection coefficients, hence the
string equations of the constrained Toda lattice hierarchies. This
proves the equivalence between type 0A (0B) MQM and IMM perturbed
by purely momentum modes or winding modes. The contour
prescriptions introduced above should be regarded as part of the
definition of the double scaling limit for the IMM.

\newsec{Connecting IMM to MQM via Tachyon Condensation}

\subsec{The array of D-instantons}

In this section we shall attempt to connect the IMM to MQM in a
more direct and intuitive way via open string tachyon
condensation. To be definite let us work with type 0B string
theory compactified on a thermal circle of radius $R=ma$, where
$m$ is an integer, $a=\pi\sqrt{2\alpha'}$ is the critical
distance. As well known \CKLM, turning on an open string tachyon
profile \eqn\senpro{T(X)=\lambda\cos (\pi X/a)} on the Euclidean
world volume of an unstable D0-brane is described by an exactly
marginal deformation in the boundary CFT. For sufficiently large
$\lambda$ one ends up with an array of alternating D- and
anti-D-instantons separated at distance $a$. From the point of
view of matrix models, this suggests that the MQM perturbed
strongly by the tachyon profile \senpro\ should be identical to
the IMM expanded near the configuration of the array of
D-instantons. Although it is not clear to us how to describe the
tachyon profile \senpro\ in the MQM in a precise way, it is very
plausible that the effect of \senpro\ is to put the MQM on a
discrete Euclidean time lattice of spacing $a$.\foot{It is well
known\KlebanovReview\ that the discretized MQM on a time lattice
of spacing $\epsilon<a$ is exactly equivalent to the continuum
theory up to a redefinition of parameters. In the critical limit
$\epsilon=a$ the discretized MQM becomes singular. We will
regularize the theory by taking $\epsilon$ slightly less than $a$.
}

Given this picture, we want to understand how the operators on
both sides are identified. When the open string tachyon is
condensed, the D0-brane effectively vanishes, so the open string
excitations are localized near the ``sites" of the time lattice
where $T(X)\sim 0$. Let us consider in the MQM a small tachyon
lump $\Phi(x)\simeq \Phi_k$ near $x=(k+\half)a$. This will shift
the zero locus of $T(X)=\lambda\cos(\pi X/a)+\Phi(X)$ to
$x_k\simeq (k+\half)a+c(-1)^k\Phi_k$, where $c=a/\pi\lambda$. So
effectively the position of the $k$th D-instanton (or
anti-instanton) is shifted by $c(-1)^k\Phi_k$. This suggests that
the positions of the D-instantons in the array in IMM should be
mapped to the open string tachyons on the corresponding sites in
the discretized MQM. Let us denote by $X_k$ the fluctuation of the
position of the $k$th cluster of instantons from the array
configuration. Then we are tempted to identify \eqn\xphiid{ X_k
\sim (-1)^k\Phi_k }

On the IMM side we shall integrate out the (complex) tachyons
which are open strings stretched between D- and anti-D-instantons,
and get an effective theory that describes only the collective
coordinates $X_k$ of the instantons. The D-instanton and
anti-D-instanton want to move toward each other so that the
tachyon can condense. Roughly one can think of this instability as
an unstable ``effective potential"\foot{ The picture we are
describing here is rather heuristic. As we will see in the next
subsection, the interaction between the eigenvalues are not simply
represented by an ordinary potential plus the repulsion through
the Vandermonde determinant. The phase transition occurring here
as the eigenvalues merge the top of the ``potential", is similar
but different from the Gross-Witten transition\refs{\GrossWitten},
the latter being well known to describe 2D gravity coupled to
$c=0$ matter\refs{\GandM,\postNewHat}. } $V(X)$ (of periodicity
$a$) felt by the eigenvalues of $X_k$'s. Similar to the picture of
$c=1$ MQM, here the eigenvalues repel each other and fill up the
``valleys" of the potential. In the large $N$ limit, the
eigenvalues are distributed in $2m$ cuts along the thermal circle.
When there are a sufficiently large number of eigenvalues, the
cuts will connect to adjacent ones and we expect a phase
transition. The transition point is where the ``energy" of the
eigenvalues reaches the top of the potential. From \xphiid\ we
expect that this is the critical point that defines the double
scaling limit! In other words, the array of D-instantons at
critical distance in IMM plays the same role as the unstable
D0-brane corresponding to the top of the tachyon potential in MQM.

\subsec{Effective matrix integral of the array}

We will expand type 0B IMM near the configuration of an array of
alternating D- and anti-D-instanton clusters at separation
$\epsilon$. It is natural to work with $\epsilon>a$ so that all
the tachyons stretched between D- and anti-D-instantons are
massive. This can be thought of as a kind of regularization.
Eventually we will be interested in the limit $\epsilon\to a$. We
will assume the radius $R=2ma/2\pi$, so there are $2m$ clusters
distributed along the thermal circle, with $N$ instantons in each
cluster. We shall write the diagonalized $X$ and $\tilde X$ as
\eqn\obarray{ X = \left( \eqalign{ X_1 & \cr & X_2+2\epsilon \cr &
~~~~~~~~\ddots \cr & ~~~~~~~~~~~~~ X_{m}+(2m-2)\epsilon }
\right),~~~~ \tilde X = \left( \eqalign{ \tilde X_1+\epsilon & \cr
& \tilde X_2+3\epsilon \cr & ~~~~~~~~\ddots \cr & ~~~~~~~~~~~~~
\tilde X_{m}+(2m-1)\epsilon } \right) } where $X_k={\rm
diag}\{x_{2k-1,\alpha}\}_{\alpha=1}^N$ and $\tilde X_k = {\rm
diag}\{x_{2k,\alpha}\}_{\alpha=1}^N$ are the collective
coordinates of each cluster of $N$ instantons.

We will take $\epsilon$ to be very close to $a$, and ignore all
the massive modes with masses of order $\sim a$, so that only the
``tachyons" as open strings stretched between adjacent clusters
are retained. Furthermore, we shall take the double scaling limit
{\sl before} taking the limit $\epsilon\to a$, so we can assume
the fluctuations $x_{j,\alpha}$ are small, since only the small
fluctuations near the top of the potential are responsible for the
universal behavior of the theory near the critical point. With
these approximations, the (unperturbed) partition function can be
written as \eqn\imparm{ Z_N = \oint \prod_{j,\alpha}
{dz_{j,\alpha}\over 2\pi i} \prod_{j=1}^{2m} {\det}_{\alpha\beta}
\left({1\over q^{-1/2} z_{j+1,\alpha} - q^{1/2}
z_{j,\beta}}\right) } where $z_{j,\alpha} =
e^{i(x_{j,\alpha}+j\epsilon)/R}$, $q=e^{ia/R}$. We can rewrite it
as \eqn\imparma{ \eqalign{ Z_N &= \int \prod_{j,\alpha}
dx_{j,\alpha}\prod_{j=1}^{2m}{\det}_{\alpha\beta} \exp\left[{-\ln
\sin ({x_{j+1,\alpha}-x_{j,\beta}+\epsilon-a\over 2R}) }\right]
\cr &= \int \prod_{j,\alpha}
dx_{j,\alpha}\prod_{j=1}^{2m}{\det}_{\alpha\beta}
\exp\left[{1\over 4R^2\sin^2({\epsilon-a\over 2R})}
(x_{j+1,\alpha}-x_{j,\beta})^2 +{\cal O}(x^3) \right] } }Let us
note that there is a zero mode corresponding to the overall shift
of $X$ and $\tilde X$, whose origin can be traced back to the
decoupled diagonal $U(1)$ in the $U(N)\times U(N)$ type 0A MQM. We
are free to fix this redundant gauge symmetry. Then we expect a
tachyonic instability for each individual $x_i$ and $\tilde x_j$
as they want to move towards each other.

Using the Itzykson-Zuber formula, we can rewrite \imparma\ in the
limit $\epsilon\to a$ as\foot{To compare with the three matrix
model, we are really taking $\epsilon$ to differ from $a$ by a
small imaginery number.} \eqn\immlim{ \eqalign{ &\int
\prod_{j=1}^{2m} dX_j d\Omega_j\, e^{{1\over(\epsilon-a)^2}
\sum_j{\rm Tr} (X_{j+1}-\Omega_jX_j\Omega_j^\dagger)^2} \cr &\sim
\int \prod_{j=1}^{2m}dX_jd\Omega_j
\delta(X_{j+1}-\Omega_jX_j\Omega_j^\dagger)\cr &= \int
\prod_{j=1}^{2m}dX_jdY_jd\Omega_j\, e^{i\sum_j{\rm
Tr}(X_{j+1}-\Omega_jX_j\Omega_j^\dagger)Y_j} } } In fact, by
integrating out all but one of the $X_j$'s, \immlim\ reduces to
the three matrix model of \KostovReview \eqn\tmma{ \int
dX_+dX_-d\Omega \,e^{i{\rm Tr}(X_+X_--qX_+\Omega
X_-\Omega^\dagger)} } with $q=e^{2\pi iR/\sqrt{2\alpha'}}$. Our
prescription is to compute correlators with imaginery $R$, and
then analytically continue to real values of
$R$($=m\sqrt{2\alpha'}$ in above).

%
%

\subsec{The discretized MQM}

Now we put the MQM on a discretized time lattice of spacing
$\epsilon$, where $\epsilon<a$. In order for the discretized MQM
to be exactly equivalent to the continuum MQM, we need to have
discretized propagator \eqn\discpra{ \int dU_{i,i+1} \exp
\left(-{\beta'\epsilon} {\rm Tr}\left[ ({U_{i,i+1}\Phi_{i+1}
U_{i,i+1}^{-1}-\Phi_{i}\over \epsilon})^2 - \half\omega'^2
(\Phi_{i+1}^2+\Phi_i^2) \right] \right) } where $dU_{i,i+1}$ is
the Haar measure over $U(N)$, and the parameters $\beta',\omega'$
are related to those of the continuum model by \eqn\disceb{
\omega'\epsilon=2\sin {\pi\epsilon\over 2a},~~~~\beta' = \beta
{\pi\epsilon/a\over \sin (\pi\epsilon/a)} } We can diagonalize
$\Phi_i$ into its eigenvalues $\lambda_{i,\alpha}$ at each step,
so the propagator simplifies to \eqn\discprb{ \int dU \exp
\left(-{\beta'\over\epsilon} \left[
\sum_\alpha(\lambda_{i+1,\alpha}^2+\lambda_{i,\alpha}^2)
-2\sum_{\alpha,\beta} \lambda_{i,\alpha}\lambda_{i+1,\beta}
U_{\alpha\beta}U_{\alpha\beta}^* - \half\omega'^2 \epsilon^2
\sum_\alpha(\lambda_{i,\alpha+1}^2+\lambda_{i,\alpha}^2) \right]
\right) } Using the Itzykson-Zuber formula \eqn\intforma{
\int_{U(N)} dU \,e^{\sum_{\alpha,\beta}
U_{\alpha\beta}U_{\alpha\beta}^*x_\alpha y_\beta} =
\prod_{k=1}^{N-1}k! {\det_{\alpha\beta} e^{x_\alpha y_\beta}\over
\Delta(x)\Delta(y)} } The propagator becomes \eqn\discprn{
\exp\left(-{\beta'\over \epsilon} (
1-\half\omega'^2\epsilon^2)\sum_\alpha
(\lambda_{i+1,\alpha}^2+\lambda_{i,\alpha}^2) \right)
{{\det}_{\alpha\beta} (e^{{2\beta'\over
\epsilon}\lambda_{i+1,\alpha}\lambda_{i,\beta} })\over
\Delta(\lambda_{i+1})\Delta(\lambda_i) } }

%
%

The partition function is given by \eqn\dmparfi{ Z_N = \int \prod
d\lambda_{i,\alpha} \prod_{i=1}^{2m}{\det}_{\alpha\beta} \exp
\left\{ -{\beta'\over \epsilon} \left[ (\lambda_{i+1,\alpha} -
\lambda_{i,\beta})^2 -\half\omega'^2\epsilon^2
(\lambda_{i+1,\alpha}^2+\lambda_{i,\beta}^2) \right] \right\} }
%
%
Note that we haven't made any approximation in above
manipulations. Under the identification \disceb, \dmparfi\ is {\sl
exactly} the same as the path integral of the continuum MQM.


In the limit $\epsilon\sim a$, \dmparfi\ is approximately
\eqn\dmparfib{ Z_N = \int \prod d\lambda_{i,\alpha}
\prod_{i=1}^{2m}{\det}_{\alpha\beta} \exp \left\{ -{\beta\over
a-\epsilon} \left[ -(\lambda_{i+1,\alpha} + \lambda_{i,\beta})^2
+{\pi^2\over 2a^2}(a-\epsilon)^2
(\lambda_{i+1,\alpha}^2+\lambda_{i,\beta}^2) \right] \right\} }
Under the identification \xphiid, \dmparfib\ approaches the path
integral of IMM expanded near the array \imparma, up to rescalings
of $\epsilon$ and $\lambda_i$'s which can be absorbed into a shift
in the chemical potential. This confirms the connection between
the discretized MQM and the IMM expanded around the D-instanton
array in the critical limit.

Similar to the manipulation of \immlim, we can rewrite \dmparfib\
as \eqn\dmqmlim{ \int \prod_{j=1}^{2m} dX_j dY_j d\Omega_j
e^{i\sum_j{\rm
Tr}(X_{j+1}-e^{-i\pi(1-\epsilon/a)}\Omega_jX_j\Omega_j^\dagger)Y_j}
} which again reduces to the three matrix model. The contour
prescription of giving $R$ a small imaginery part so that
$|q|=|e^{2\pi iR}|<1$, corresponds to taking $\epsilon-a$ to be a
small imaginery number.

%
%

\newsec{On SD-branes and ZZ Boundary States}

It was pointed out in \MSY\ that sD-branes ($\lambda=1/2$ s-brane)
can be described as an array of D-instantons along Euclidean time
separated at the critical distance. In the previous section we
have argued that sD-branes play the same role in the IMM as
D0-branes in MQM. In this section we study the closed string
fields sourced by sD-branes from IMM. We find agreement with
calculations from ZZ boundary states. This is very much in the
same spirit as the calculation of D-brane decay into closed
strings from MQM \KMSdecay\ as opposed to the boundary state
approach. In fact, we will reproduce the $(1,1)$ ZZ boundary
states from IMM in a very simple manner.

\subsec{sD-brane in $c=1$ string theory}

In this subsection we study the closed string fields in spacetime
dual to the array configuration in the IMM for $c=1$ string. The
configuration of an array of D-instantons separated at distance
$a=2\pi\sqrt{\alpha'}$ ($2\pi R=ma$) is described by \eqn\caarray{
X = \left( \eqalign{ {1\over 2}a  & \cr &~ {3\over2}a \cr &
~~~~~~~~\ddots \cr & ~~~~~~~~~~~~~ (m-\half)a  } \right) }

The array configuration can be expressed in a gauge invariant way
as \eqn\arraygi{ {\rm Tr} U^k = \left\{ \eqalign{& (-1)^{k/m}{m
},~~~ {\rm if}~m|k, \cr &0,~~~{\rm otherwise}.} \right. } for all
$k\in{\bf Z}$, where $U=e^{iX/R}$. The operator ${\rm Tr}U^k$ is
mapped to vertex operator\KKK \eqn\truvert{ {\rm Tr}U^k
\leftrightarrow {1\over
(q^{k/2}-q^{-k/2})}{\Gamma(|p|)\over\Gamma(-|p|)} \int
e^{ipX+(2-|p|)\phi},~~~~ p=k/R } Therefore \arraygi\ corresponds
to the condensation of momentum modes \eqn\momcond{\langle
V_{l}\rangle = \langle e^{ilX+(2-|l|)\phi} \rangle =
(q^{ml/2}-q^{-ml/2}){\Gamma(-|l|)\over\Gamma(|l|)}(-1)^l
,~~~l\in{\bf Z}. } Formally $\Gamma(-|l|)={(-1)^{l}\over
|l|!}\Gamma(0)$ is divergent, and since $q=e^{2\pi i/R}=e^{2\pi
i/m}$, $(q^{ml/2}-q^{-ml/2})$ is zero. As discussed earlier (also
as in \UpsideDown), to make the IMM well-defined, we should add a
small imaginery part to $R$, say $R\to R(1-i\epsilon)$ so that
$|q|<1$, $$ q^{ml/2}-q^{-ml/2}\to (-1)^l (-2\pi l\epsilon) $$ At
the same time, we have
$$\Gamma(-|l|)\to \Gamma(-{|l|\over 1-i\epsilon}) = {(-1)^l\over
|l|!}{1\over (-i\epsilon) |l|}$$ The factors of $\epsilon$ cancel,
and we end up with spacetime tachyon profile \eqn\tachpro{
\eqalign{ T(X,\phi) &= \sum_l {1\over l}\langle V_l\rangle
\cos(lX)e^{l\phi} \cr &\sim 2\sum_{l=0}^\infty {(-1)^l\over
(l!)^2}\mu^{l/2}\cos (lX) e^{l\phi} \cr &= J_0(2\mu^{1/4}
e^{\phi+iX\over 2}) + J_0(2\mu^{1/4} e^{\phi-iX\over 2}) } } This
result precisely agrees with the closed string fields sourced by
sD-branes computed from ZZ boundary states \tadashi.

In fact, we can consider a much simpler case, - a single
D-instanton sitting at $X=0$. This is described by ${\rm Tr}
U^k=1$ for all $k\in{\bf Z}$ in the IMM. As above, it corresponds
to a condensation of closed string momentum modes \eqn\sdinst{
\langle V_{p=k/R}\rangle = (q^{k/2}-q^{-k/2}) {\Gamma(-|p|)\over
\Gamma(|p|)} = 2i\sin(\pi p){\Gamma(-|p|)\over \Gamma(|p|)} } This
is nothing but the ZZ boundary state (times the Dirichlet boundary
state in Euclidean time direction), up to a normalization factor.
The agreement between IMM calculation and ZZ boundary state is
very reminiscent to the calculation of \KMSdecay.

\subsec{sD-brane in type 0B string theory}

An sD-brane in type 0B theory is described by an array of
alternating D- and anti-D-instantons separated at critical
distance $a=\pi\sqrt{2\alpha'}$. In this subsection we will set
$\alpha'=2$ for convenience. The radius is then $R=2ma/2\pi=2m$,
and we have $q=e^{ia/R}=e^{\pi i/m}$. The D-instantons are located
$X={1\over 2}a,{5\over 2}a,\cdots, (2m-{3\over 2})a$, whereas the
anti-D-instantons are located at $\tilde X={3\over 2}a, {7\over 2}
a,\cdots,(2m-{1\over 2})a$. They can be described in the following
gauge invariant way \eqn\arraygiob{ {\rm Tr} U^k = \left\{
\eqalign{& e^{i{k\pi\over 2m}}{m },~~~ {\rm if}~m|k, \cr
&0,~~~{\rm otherwise}.} \right.~~~~~~~~~ {\rm Tr} \tilde U^k =
\left\{ \eqalign{& e^{-i{k\pi\over 2m}}{m},~~~ {\rm if}~m|k, \cr
&0,~~~{\rm otherwise}.} \right. } From \idsa, they are mapped to
NS-NS and R-R vertex operators \eqn\trutovertst{ \eqalign{ &
{1\over 2}({\rm Tr} U^k + {\rm Tr} \tilde U^k) \leftrightarrow
{1\over (q^{k/2}-q^{-k/2})}{\Gamma(|p|)\over\Gamma(-|p|)} \int
e^{ipX + (1-|p|)\phi} \cr & {1\over 2}({\rm Tr} U^k - {\rm Tr}
\tilde U^k) \leftrightarrow {1\over
(q^{k/2}+q^{-k/2})}{\Gamma(\half+|p|)\over\Gamma(\half-|p|)} \int
e^{ipX + (1-|p|)\phi} S\bar S } } where $p=k/R$, and $S,\bar S$
are spin fields. On RHS of \trutovertst\ we have included the
corresponding leg factors. \arraygiob\ corresponds to condensation
of closed string modes \eqn\condsu{ \eqalign{ \langle
V_{NS,p=l}\rangle &=
(q^{ml}-q^{-ml}){\Gamma(-|l|)\over\Gamma(|l|)} (-1)^l\cr &\to
l{(-1)^l\over (|l|!)^2}, \cr \langle V_{R,p=l+\half}\rangle &=
i(q^{m(l+1/2)}+q^{-m(l+1/2)}){\Gamma(\half-|l+\half|)\over\Gamma(\half+|l+\half|)}
(-1)^l\cr &\to {l+\half\over|l+\half|}\cdot{(-1)^l\over
[(|l+\half|-\half)!]^2} } } where $l\in{\bf Z}$, and we have again
used the prescription $R\to R(1-i\epsilon)$ to regularize the
singular terms. The NS-NS and R-R scalars in spacetime are
\eqn\tachrrpro{ \eqalign{ T(X,\phi) &\sim 2\sum_{l=0}^\infty
{(-1)^l\over (l!)^2} (2\mu)^l\cos (lX) e^{l\phi} \cr &=
J_0(2\sqrt{2\mu}e^{\phi+iX\over2}) +
J_0(2\sqrt{2\mu}e^{\phi-iX\over 2}), \cr V(X,\phi) &\sim
2\sum_{l=0}^\infty {(-1)^l\over (l!)^2}{1\over
l+\half}(2\mu)^{l+\half} \sin((l+\half)X) e^{(l+\half)\phi}.
 } } In particular, we find using \tachrrpro\ that (with proper normalization)
\eqn\schar{ \eqalign{ \int_{-\infty}^\infty d\phi \partial_X
V(X,\phi) = \int_{-\infty}^\infty d\phi\, {1\over
4}\left[\sqrt{2\mu} e^{\phi+iX\over 2} J_0(2\sqrt{2\mu}
e^{\phi+iX\over2}) \right.\cr\left.+ \sqrt{2\mu} e^{\phi-iX\over2}
J_0(2\sqrt{2\mu} e^{\phi-iX\over 2})\right] = {1\over2} }} This is
the conserved s-charge of the sD-brane\MSY.

Extending \tadashi, we can compute the closed string fields in the
weak coupling region from the ZZ boundary state in super-Liouville
theory. Our convention is $b=1,Q=2$, so that $\hat c_L =1+2Q^2=9$.
The disk 1-point function for the tachyon $T(P)$ and RR scalar
$V(P)$ are \refs{\ZZsupera,\ZZsuperb,\NewHat} \eqn\nsrzzsl{
\eqalign{ & \psi_{NS}(P) = -i\sqrt{2\pi} \sinh(\pi P)
{\Gamma(iP)\over \Gamma(-iP)} (2\mu)^{-iP}, \cr & \psi_{R}(P) =
\sqrt{2\pi} \cosh(\pi P) {\Gamma(\half+iP)\over \Gamma(\half-iP)}
(2\mu)^{-iP}. } } The closed string fields sourced by the sD-brane
are (in momentum space) \eqn\csfsdb{ \eqalign{ T(P,t) &= {1\over
4\pi E} \psi_{NS}(P)\sum_{n\geq 0}e^{-(n+\half)aE}
(e^{-iEt}+e^{iEt}) \cr & = {1\over 4\pi E}\psi_{NS}(P) {\cos
Et\over\sinh(aE/2)}, \cr V(P,t) &= {1\over 4\pi E}
\psi_{R}(P)\sum_{n\geq 0} (-1)^n e^{-(n+\half)aE}
(e^{-iEt}-e^{iEt}) \cr&= {-i\over 4\pi E}\psi_{R}(P) {\sin
Et\over\cosh(aE/2)}, } } where $E=|P|$. Translated into Liouville
coordinates, \eqn\nsrob{ \eqalign{ T(\phi,t) &\sim
-i\int_{-\infty}^\infty dP e^{-iP\phi}{\cos{Pt}\over P}
{\Gamma(iP)\over \Gamma(-iP)} (2\mu)^{-iP} \cr &= J_0(2\sqrt{2\mu}
e^{\phi+t\over2}) + J_0(2\sqrt{2\mu} e^{\phi-t\over 2}), \cr
V(\phi,t) &\sim -i\int_{-\infty}^\infty dP
e^{-iP\phi}{\sin{Pt}\over P} {\Gamma(\half+iP)\over
\Gamma(\half-iP)} (2\mu)^{-iP} \cr &= 2\sum_{l=0}^\infty
{(-1)^l\over (l!)^2}{1\over l+\half}(2\mu)^{l+\half}
\sinh((l+\half)t) e^{(l+\half)\phi} } } which precisely agree with
\tachrrpro.

Again, we can consider the simpler case of a single D-instanton at
$X=0$. This is described in IMM as ${\rm Tr}U^k=1,{\rm Tr}\tilde
U^k=0$ for all $k\in{\bf Z}$. It then follows from \trutovertst\
that \eqn\zzsu{ \eqalign{ &\langle V_{NS,p=k/R}\rangle =
(q^{k/2}-q^{-k/2}){\Gamma(-|p|)\over \Gamma(|p|)} = 2i\sin(\pi p)
{\Gamma(-|p|)\over \Gamma(|p|)} \cr &\langle V_{RR,p=k/R}\rangle =
(q^{k/2}+q^{-k/2}){\Gamma(\half-|p|)\over\Gamma(\half+|p|)} =
2\cos(\pi p) {\Gamma(\half-|p|)\over\Gamma(\half+|p|)} } } which
reproduce up to a normalization factor the ZZ boundary state of
super-Liouville theory \nsrzzsl. This is also a nontrivial check
of the dictionary \idsa.

\newsec{On Black Holes in Type 0 String Theory}
We have shown that type 0A and 0B string theories have the
integrable structure of Toda lattice hierarchy. This has enabled
us to prove the T-duality between type 0A and type 0B MQM at least
for perturbations by purely momentum modes or winding modes. The
matrix model deformed by modes of winding number $\pm1$ is
equivalent (in the $\mu\to0$ limit) to the sine-Liouville theory
(${\cal N}=2$ Liouville theory in type 0 case), which is believed
to be dual to string theory in the Euclidean 2D black hole
\refs{\KKK,\fermiBH,\Kutasov}. The exact free energy of the
deformed type 0A and 0B MQM can be computed by solving the Hirota
differential equations.

Let us first consider the case of type 0B MQM. The winding mode
perturbations, just as in $c=1$ string, generate the Toda
integrable flow. The unperturbed free energy of type 0B MQM is
perturbatively the same as that of $c=1$ string, up to a
redefinition $\alpha'\to 2\alpha'$ and an overall factor of 2
(coming from doubling the fermi sea). The method used in \KKK\ to
compute the free energy perturbed by $\lambda_{\pm}\equiv
t_{\pm1}$ can be directly applied to type 0B case. The overall
factor of 2 in the unperturbed free energy is very important,
since the Hirota differential equations are nonlinear. In the
limit of large $\lambda$ and $\mu=0$, the free energy is now (in
units $\alpha'=1/2$) \eqn\freeexpan{ {\cal F}(\lambda,\mu=0,R) = -
(1-R)^2(2R-1)^{R\over 1-R} \lambda^{2\over 1-R} - {R+R^{-1}\over
12 (1-R)}\ln (\lambda\sqrt{2R-1})+\cdots } where we have exhibited
the genus 0 and genus 1 terms in the expansion. The asymptotic
radius of the Euclidean black hole in type 0B string is
$R=\sqrt{\alpha'/2}=1/2$. Expanding near this radius, we
have\foot{We are grateful to A. Adams for discussions on this
point.} \eqn\freenbh{ {\cal F} = -2\pi(R-{1\over2})M - {5\over 12}
\ln M+\cdots } where $M\propto \lambda^{2\over 1-R}$. At $R=1/2$,
$M$ is expected to be the mass of the black hole. In fact, the
effective string coupling in ${\cal N}=2$ Liouville theory is
$g_s\lambda^{-2}$, which is the combination invariant under
shifting the Liouville coordinate. The mass of the black hole goes
like $M\sim 1/g_s^2$, therefore must be proportional to
$\lambda^4$. This agrees with the expectation from the matrix
model result \freenbh.

The density of states in the black hole background typically has
Hagedorn growth behavior \eqn\haged{ \rho(M) \sim M^{s_1}
e^{\beta_H M} } Comparing with the free energy near $T_H$
\freenbh, we find \eqn\sone{ s_1+1=-{R+R^{-1}\over 24}=-{5\over
12} } This can be compared to the $c=1$ string case, where the
asymptotic radius of the 2D black hole is $R=3/2$, and one finds
$s_1+1=-(R+R^{-1})/48=-13/288$ \KKK.

One can study Euclidean black holes in type 0A theory in a similar
way. The uncharged black holes should be described by a background
with the lowest NS-NS winding modes condensed. As in \trutovertst,
the perturbation of these winding modes are represented in type 0A
MQM as deformations by the operators ${\rm Tr}\Omega+{\rm
Tr}\tilde\Omega$ and ${\rm Tr}\Omega^{-1}+{\rm
Tr}\tilde\Omega^{-1}$. Strictly speaking, the coefficients of
these perturbations are not the same as the time variables that
generate the integrable flow, but essentially related to them
through nonlinear bosonization maps, as we have seen in section
3.4. It then appears much more complicated to solve the
differential constraints on the free energy exactly.

However, if we are only interested in the perturbative expansion
of the free energy, the calculation is greatly simplified. In the
T-dual type 0B picture, perturbatively the two sides of the fermi
sea decouple. The deformation by NS-NS winding modes can be
treated as independent perturbations in the two decoupled sectors.
Each sector is perturbatively the same as $c=1$ string, up to a
redefinition of $\alpha'$. In the end, the free energy of type 0A
theory deformed by NS-NS $\pm1$ winding modes is simply obtained
from the solution of \KKK\ for $c=1$ string with a replacement
$\alpha'\to\alpha'/2$, and a factor of 2 coming from the two
sectors\foot{Unlike the case of type 0B theory, this factor of 2
trivially multiplies the answer of the perturbed free energy.}. In
units with $\alpha'=2$, the answer is \eqn\freeexpanoa{ {\cal
F}(\lambda,\mu=0,R) = - {1\over 2} (2-R)^2(R-1)^{R\over 2-R}
\lambda^{4\over 2-R} - {R+R^{-1}\over 6 (2-R)}\ln
(\lambda\sqrt{R-1})+\cdots } Again, near the black hole asymptotic
radius $R=\sqrt{\alpha'/2}=1$, we have \eqn\freenbhoa{ {\cal F} =
-2\pi(R-1)M - {1\over 12} \ln M+\cdots } where $M\propto
\lambda^{4\over 2-R}\simeq\lambda^4$. The growth of the density
states is of the form \haged\ with the exponent $s_1$ given by
\eqn\soneoa{ s_1+1=-{R+R^{-1}\over 24}=-{1\over 12}. }

It would be interesting to reproduce the exponents \sone, \soneoa\
from 1-loop calculations in ${\cal N}=2$ Liouville theory. One can
also consider the sectors with nonzero net D0-brane charge $q$
perturbed by the lowest winding modes, which should lead to
charged nonextremal black holes.

\bigskip

\centerline{\bf Acknowledgements}

I would like to thank J. Karczmarek and A. Strominger for
collaboration at the initial stage of this work, as well as many
helpful conversations. I'm also grateful to A. Adams, M. A.
Belabbas, S. Gukov, P.-M. Ho, D. Jafferis, G. Jones, I. Kostov, A.
Maloney, J. Marsano, S. Minwalla, A. Neitzke, and T. Takayanagi
for very useful discussions.

\listrefs

\end